\newcount\mgnf\newcount\tipi\newcount\tipoformule\newcount\greco

\tipi=2 
\tipoformule=0

\global\newcount\numsec
\global\newcount\numfor
\global\newcount\numtheo
\global\advance\numtheo by 1

\def\senondefinito#1{\expandafter\ifx\csname#1\endcsname\relax}

\def\SIA #1,#2,#3 {\senondefinito{#1#2}%
\expandafter\xdef\csname #1#2\endcsname{#3}\else
\write16{???? ma #1,#2 e' gia' stato definito !!!!} \fi}

\def\etichetta(#1){(\veroparagrafo.\veraformula)%
\SIA e,#1,(\veroparagrafo.\veraformula) %
\global\advance\numfor by 1%
\write15{\string\FU (#1){\equ(#1)}}%
\write16{ EQ #1 ==> \equ(#1) }}

\def\letichetta(#1){\veroparagrafo.\verotheo
\SIA e,#1,{\veroparagrafo.\verotheo}
\global\advance\numtheo by 1
\write15{\string\FU (#1){\equ(#1)}}
\write16{ Sta \equ(#1) == #1 }}

\def\tetichetta(#1){\veroparagrafo.\veraformula 
\SIA e,#1,{(\veroparagrafo.\veraformula)}
\global\advance\numfor by 1
\write15{\string\FU (#1){\equ(#1)}}
\write16{ tag #1 ==> \equ(#1)}}

\def\FU(#1)#2{\SIA fu,#1,#2 }

\def\etichettaa(#1){({\rm A}.\veraformula)%
\SIA e,#1,(A.\veraformula) %
\global\advance\numfor by 1%
\write15{\string\FU (#1){\equ(#1)}}%
}

\def\etichettab(#1){({\rm B}.\veraformula)%
\SIA e,#1,(B.\veraformula) %
\global\advance\numfor by 1%
\write15{\string\FU (#1){\equ(#1)}}%
}
\def\etichettac(#1){({\rm C}.\veraformula)%
\SIA e,#1,(C.\veraformula) %
\global\advance\numfor by 1%
\write15{\string\FU (#1){\equ(#1)}}%
}
\def\etichettad(#1){({\rm D}.\veraformula)%
\SIA e,#1,(D.\veraformula) %
\global\advance\numfor by 1%
\write15{\string\FU (#1){\equ(#1)}}%
}

\def\BOZZA{
\def\alato(##1){%
{\rlap{\kern-\hsize\kern-1.4truecm{$\scriptstyle##1$}}}}%
\def\aolado(##1){%
{
{
\rlap{\kern-1.4truecm{$\scriptstyle##1$}}}}}
}

\def\alato(#1){}
\def\aolado(#1){}

\def\veroparagrafo{\number\numsec}
\def\veraformula{\number\numfor}
\def\verotheo{\number\numtheo}

\def\Eq(#1){\eqno{\etichetta(#1)\alato(#1)}}
\def\eq(#1){\etichetta(#1)\alato(#1)}
\def\leq(#1){\leqno{\aolado(#1)\etichetta(#1)}}
\def\Eqa(#1){\eqno{\etichettaa(#1)\alato(#1)}}
\def\eqa(#1){\etichettaa(#1)\alato(#1)}
\def\Eqb(#1){\eqno{\etichettab(#1)\alato(#1)}}
\def\eqb(#1){\etichettab(#1)\alato(#1)}
\def\Eqc(#1){\eqno{\etichettac(#1)\alato(#1)}}
\def\eqc(#1){\etichettac(#1)\alato(#1)}
\def\Eqd(#1){\eqno{\etichettad(#1)\alato(#1)}}
\def\eqd(#1){\etichettad(#1)\alato(#1)}

\def\eqv(#1){\senondefinito{fu#1}$\clubsuit$#1
\write16{#1 non e' (ancora) definito}%
\else\csname fu#1\endcsname\fi}
\def\equ(#1){\senondefinito{e#1}\eqv(#1)\else\csname e#1\endcsname\fi}

\def\Lemma(#1){\aolado(#1)Lemma \letichetta(#1)}%
\def\Theorem(#1){{\aolado(#1)Theorem \letichetta(#1)}}%
\def\Proposition(#1){\aolado(#1){Proposition \letichetta(#1)}}%
\def\Corollary(#1){{\aolado(#1)Corollary \letichetta(#1)}}%
\def\Remark(#1){{\noindent\aolado(#1){\bf Remark \letichetta(#1).}}}%
\def\Definition(#1){{\noindent\aolado(#1){\bf Definition
\letichetta(#1)$\!\!$\hskip-1.6truemm}}}
\def\Example(#1){\aolado(#1) Example \letichetta(#1)$\!\!$\hskip-1.6truemm}

\def\include#1{
\openin13=#1.aux \ifeof13 \relax \else
\input #1.aux \closein13 \fi}

\openin14=\jobname.aux \ifeof14 \relax \else
\input \jobname.aux \closein14 \fi
\openout15=\jobname.aux

{\count255=\time\divide\count255 by 60 \xdef\hourmin{\number\count255}
\multiply\count255 by-60\advance\count255 by\time
\xdef\hourmin{\hourmin:\ifnum\count255<10 0\fi\the\count255}}

\def\oramin{\hourmin }

\def\data{\number\day/\ifcase\month\or january \or february \or march
\or april
\or may \or june \or july \or august \or september
\or october \or november \or december \fi/\number\year;\ \oramin}

\newcount\pgn \pgn=1
\def\foglio{\number\numsec:\number\pgn
\global\advance\pgn by 1}
\def\foglioa{A\number\numsec:\number\pgn
\global\advance\pgn by 1}

\footline={\rlap{\hbox{\copy200}}\hss\tenrm\folio\hss}
\def\TIPIO{
\font\setterm=amr7 
\def \settepunti{\def\rm{\fam0\setterm}
\textfont0=\setterm 
\normalbaselineskip=9pt\normalbaselines\rm }\let\nota=\settepunti}

\def\TIPITOT{
\font\twelverm=cmr12
\font\twelvei=cmmi12
\font\twelvesy=cmsy10 scaled\magstep1
\font\twelveex=cmex10 scaled\magstep1
\font\twelveit=cmti12
\font\twelvett=cmtt12
\font\twelvebf=cmbx12
\font\twelvesl=cmsl12
\font\ninerm=cmr9
\font\ninesy=cmsy9
\font\eightrm=cmr8
\font\eighti=cmmi8
\font\eightsy=cmsy8
\font\eightbf=cmbx8
\font\eighttt=cmtt8
\font\eightsl=cmsl8
\font\eightit=cmti8
\font\sixrm=cmr6
\font\sixbf=cmbx6
\font\sixi=cmmi6
\font\sixsy=cmsy6
\font\twelvetruecmr=cmr10 scaled\magstep1
\font\twelvetruecmsy=cmsy10 scaled\magstep1
\font\tentruecmr=cmr10
\font\tentruecmsy=cmsy10
\font\eighttruecmr=cmr8
\font\eighttruecmsy=cmsy8
\font\seventruecmr=cmr7
\font\seventruecmsy=cmsy7
\font\sixtruecmr=cmr6
\font\sixtruecmsy=cmsy6
\font\fivetruecmr=cmr5
\font\fivetruecmsy=cmsy5
\textfont\truecmr=\tentruecmr
\scriptfont\truecmr=\seventruecmr
\scriptscriptfont\truecmr=\fivetruecmr
\textfont\truecmsy=\tentruecmsy
\scriptfont\truecmsy=\seventruecmsy
\scriptscriptfont\truecmr=\fivetruecmr
\scriptscriptfont\truecmsy=\fivetruecmsy
\def \eightpoint{\def\rm{\fam0\eightrm}
\textfont0=\eightrm \scriptfont0=\sixrm \scriptscriptfont0=\fiverm
\textfont1=\eighti \scriptfont1=\sixi \scriptscriptfont1=\fivei
\textfont2=\eightsy \scriptfont2=\sixsy \scriptscriptfont2=\fivesy
\textfont3=\tenex \scriptfont3=\tenex \scriptscriptfont3=\tenex
\textfont\itfam=\eightit \def\it{\fam\itfam\eightit}%
\textfont\slfam=\eightsl \def\sl{\fam\slfam\eightsl}%
\textfont\ttfam=\eighttt \def\tt{\fam\ttfam\eighttt}%
\textfont\bffam=\eightbf \scriptfont\bffam=\sixbf
\scriptscriptfont\bffam=\fivebf \def\bf{\fam\bffam\eightbf}%
\tt \ttglue=.5em plus.25em minus.15em
\setbox\strutbox=\hbox{\vrule height7pt depth2pt width0pt}%
\normalbaselineskip=9pt
\let\sc=\sixrm \let\big=\eightbig \normalbaselines\rm
\textfont\truecmr=\eighttruecmr
\scriptfont\truecmr=\sixtruecmr
\scriptscriptfont\truecmr=\fivetruecmr
\textfont\truecmsy=\eighttruecmsy
\scriptfont\truecmsy=\sixtruecmsy }\let\nota=\eightpoint}

\newfam\msbfam 
\newfam\truecmr 
\newfam\truecmsy 
\newskip\ttglue
\ifnum\tipi=0\TIPIO \else\ifnum\tipi=1 \TIPI\else \TIPITOT\fi\fi

\def\a{\alpha}
\def\b{\beta}

\def\e{\epsilon}

\def\f{\phi}

\def\k{\kappa}

\def\r{\rho}
\def\s{\sigma}

\def\L{\Lambda}

\def\h{\eta}

\def\E{{I\kern-.25em{E}}}
\def\N{{I\kern-.25em{N}}}
\def\M{{I\kern-.25em{M}}}
\def\R{{I\kern-.25em{R}}}
\def\Z{{Z\kern-.425em{Z}}}
\def\1{{1\kern-.25em\hbox{\rm I}}}
\def\eu{{1\kern-.25em\hbox{\sm I}}}

\def\C{{I\kern-.64em{C}}}
\def\P{{I\kern-.25em{P}}}
\def\eop{{ \vrule height7pt width7pt depth0pt}\par\bigskip}



\def\ZZ{{\cal Z}}

\def\chap #1#2{\line{\ch #1\hfill}\numsec=#2\numfor=1}

\def\sqr#1#2{{\vcenter{\vbox{\hrule height.#2pt
\hbox{\vrule width.#2pt height#1pt \kern#1pt
\vrule width.#2pt}\hrule height.#2pt}}}}


\newcount\foot
\foot=1
\def\note#1{\footnote{${}^{\number\foot}$}{\ftn #1}\advance\foot by 1}

\def\frac#1#2{{#1\over #2}}

\def\text#1{\quad{\hbox{#1}}\quad}

\def\thanks{\noindent{\bf Aknowledgements: }}



\font\ch=cmbx12
\font\ftn=cmr8

\font\it=cmti10
\font\bf=cmbx10
\font\sm=cmr7

%
\catcode`\X=12\catcode`\@=11
\def\n@wcount{\alloc@0\count\countdef\insc@unt}
\def\n@wwrite{\alloc@7\write\chardef\sixt@@n}
\def\n@wread{\alloc@6\read\chardef\sixt@@n}
\def\crossrefs#1{\ifx\alltgs#1\let\tr@ce=\alltgs\else\def\tr@ce{#1,}\fi
\n@wwrite\cit@tionsout\openout\cit@tionsout=\jobname.cit
\write\cit@tionsout{\tr@ce}\expandafter\setfl@gs\tr@ce,}
\def\setfl@gs#1,{\def\@{#1}\ifx\@\empty\let\next=\relax
\else\let\next=\setfl@gs\expandafter\xdef
\csname#1tr@cetrue\endcsname{}\fi\next}
\newcount\sectno\sectno=0\newcount\subsectno\subsectno=0\def\r@s@t{\relax}
\def\resetall{\global\advance\sectno by 1\subsectno=0
\gdef\firstpart{\number\sectno}\r@s@t}
\def\resetsub{\global\advance\subsectno by 1
\gdef\firstpart{\number\sectno.\number\subsectno}\r@s@t}
\def\v@idline{\par}\def\firstpart{\number\sectno}
\def\l@c@l#1X{\firstpart.#1}\def\gl@b@l#1X{#1}\def\t@d@l#1X{{}}
\def\m@ketag#1#2{\expandafter\n@wcount\csname#2tagno\endcsname
\csname#2tagno\endcsname=0\let\tail=\alltgs\xdef\alltgs{\tail#2,}%
\ifx#1\l@c@l\let\tail=\r@s@t\xdef\r@s@t{\csname#2tagno\endcsname=0\tail}\fi
\expandafter\gdef\csname#2cite\endcsname##1{\expandafter
\ifx\csname#2tag##1\endcsname\relax?\else{\rm\csname#2tag##1\endcsname}\fi
\expandafter\ifx\csname#2tr@cetrue\endcsname\relax\else
\write\cit@tionsout{#2tag ##1 cited on page \folio.}\fi}%
\expandafter\gdef\csname#2page\endcsname##1{\expandafter
\ifx\csname#2page##1\endcsname\relax?\else\csname#2page##1\endcsname\fi
\expandafter\ifx\csname#2tr@cetrue\endcsname\relax\else
\write\cit@tionsout{#2tag ##1 cited on page \folio.}\fi}%
\expandafter\gdef\csname#2tag\endcsname##1{\global\advance
\csname#2tagno\endcsname by 1%
\expandafter\ifx\csname#2check##1\endcsname\relax\else%
\fi
\expandafter\xdef\csname#2check##1\endcsname{}%
\expandafter\xdef\csname#2tag##1\endcsname
{#1\number\csname#2tagno\endcsnameX}%
\write\t@gsout{#2tag ##1 assigned number \csname#2tag##1\endcsname\space
on page \number\count0.}%
\csname#2tag##1\endcsname}}%
\def\m@kecs #1tag #2 assigned number #3 on page #4.%
{\expandafter\gdef\csname#1tag#2\endcsname{#3}
\expandafter\gdef\csname#1page#2\endcsname{#4}}
\def\re@der{\ifeof\t@gsin\let\next=\relax\else
\read\t@gsin to\t@gline\ifx\t@gline\v@idline\else
\expandafter\m@kecs \t@gline\fi\let \next=\re@der\fi\next}
\def\t@gs#1{\def\alltgs{}\m@ketag#1e\m@ketag#1s\m@ketag\t@d@l p
\m@ketag\gl@b@l r \n@wread\t@gsin\openin\t@gsin=\jobname.tgs \re@der
\closein\t@gsin\n@wwrite\t@gsout\openout\t@gsout=\jobname.tgs }
\outer\def\localtags{\t@gs\l@c@l}
\outer\def\globaltags{\t@gs\gl@b@l}
\outer\def\newlocaltag#1{\m@ketag\l@c@l{#1}}
\outer\def\newglobaltag#1{\m@ketag\gl@b@l{#1}}

\def\t@gsoff#1,{\def\@{#1}\ifx\@\empty\let\next=\relax\else\let\next=\t@gsoff
\expandafter\gdef\csname#1cite\endcsname{\relax}
\expandafter\gdef\csname#1page\endcsname##1{?}
\expandafter\gdef\csname#1tag\endcsname{\relax}\fi\next}
\def\verbatimtags{\let\ift@gs=\iffalse\ifx\alltgs\relax\else
\expandafter\t@gsoff\alltgs,\fi}
\catcode`\X=11 \catcode`\@=\active
\localtags
%
\global\newcount\numpunt
\hoffset=0.cm
\baselineskip=14pt
\parindent=12pt
\lineskip=4pt\lineskiplimit=0.1pt
\parskip=0.1pt plus1pt

\hyphenation{small}
\def\IR{{\rm I\kern -1.6pt{\rm R}}}
\def\IN{{\rm I\kern -1.6pt{\rm N}}}
\def\IH{{\rm I\kern -1.6pt{\rm H}}}
\def\IP{{\rm I\kern -1.6pt{\rm P}}}
\def\ZZ{{\rm Z\kern -4.0pt{\rm Z}}}
\def\IC{{\rm I\kern -6.0pt{\rm C}}}

\def\F{{\cal F}}
\def\ro{\rho_1}
\def\rt{\rho_2}

\def\La{\Lambda}
\def\vol{|\Lambda|}

\let\hat=\widehat
\let\tilde=\widetilde

\def\sqr#1#2{{\vcenter{\vbox{\hrule height.#2pt
\hbox{\vrule width.#2pt height #1pt \kern#1pt
\vrule width.#2pt}
\hrule height.#2pt}}}}

\def\pmb#1{\setbox0=\hbox{$#1$}%
\kern-.025em\copy0\kern-\wd0
\kern.05em\copy0\kern-\wd0
\kern-.025em\raise.0433em\box0 }

\def\pmbb#1{\setbox0=\hbox{$\scriptstyle#1$}%
\kern-.025em\copy0\kern-\wd0
\kern.05em\copy0\kern-\wd0
\kern-.025em\raise.0433em\box0 }
\count0=1
\catcode`\@=11
\centerline{\bf Free Energy Minimizers for a Two--Species
Model}
\centerline{\bf with Segregation and Liquid-Vapor Transition.}
\vskip .5truecm
\centerline{by}
\vskip.5cm
\centerline{E. A. Carlen\footnote{$^{(1)}$}{\eightrm School of
Mathematics, Georgia
Institute of Technology, Atlanta, GA 30332--0160},
\hskip.2cm
M. C. Carvalho \footnote{$^{(2)}$}
{\eightrm Department of Mathematics and GFM, University of Lisbon,
Av. Prof. Gama Pinto no 2 1649-003 Lisbon Portugal },
\hskip.2cm
R. Esposito\footnote{$^{(3)}$}
{\eightrm Centro Linceo ''Beniamino Segre'', and Dip. di
Matematica, Universit\`a di L'Aquila, Coppito, 67100 AQ, Italy},
\hskip .2cm
J. L. Lebowitz\footnote {$^{(4)}$}
{\eightrm Departments of Mathematics and
Physics, Rutgers University, New Brunswick, NJ
08903, USA}
\hskip.1cm and \hskip.1cm
R. Marra\footnote
{$^{(5)}$}
{\eightrm Dipartimento di Fisica and Unit\`a INFM, Universit\`a di
Roma Tor Vergata, 00133 Roma, Italy.}
}
\bigskip
\bigskip {\baselineskip = 12pt\narrower{\noindent {\bf Abstract }\/}

We study the coexistence of phases in a two--species model  whose free
energy is given
by the scaling limit of  a system with
long range interactions (Kac potentials) which are attractive between 
particles of the same
species and
repulsive between different species.

\bigskip}
\noindent{\bf Key words:} phase transition, segregation, convexity, 
rearrangement inequalities

\vskip .5 true in

\chap {1. Introduction. } 1

\bigskip

The rigorous {\it ab initio} derivation of the full phase diagram
of even the
simplest realistic model of a physical system, is still far beyond our
mathematical grasp.  This is so even when we restrict ourselves to simple 
classical one component system, e.g., point particles with Lennard-Jones
type pair interactions.  Only the low density--high temperature
properties of such a system can be fully derived from statistical
mechanics: there one can prove analytic behavior of the free
energy as a function of the fugacity (or density) and temperature.  Beyond
that we have almost no means  to prove the commonly observed facts of
the phase diagram such as the Gibbs phase rule, the smoothness of the
boundaries between the phases, etc. [\rcite{Ru}]

To overcome this deficiency one can consider even more simplified models
such as lattice systems.  There, thanks to the Pirogov-Sinai [\rcite{PS}], [\rcite{Za}]
theory, some exactly solvable models, inequalities, etc., one has much
better control over the low temperature region of the phase diagram
[\rcite{Ru}], [\rcite{Si}], [\rcite{Ge}], [\rcite{Sim}].  Alternatively,
one can consider, following the pioneering work of van der Waals
[\rcite{VW}], [\rcite{LS}],  [\rcite{Ro}], [\rcite{db}], mean-field-theories (MFT) yielding approximate
equations for state or free energies which indeed exhibit most of the
qualitative and many quantitative features of real world phase diagrams.
These results were initially obtained in a heuristic manner and had to be
supplemented by additional rules, e.g.\ the Gibbs double tangent or Maxwell
equal area rule, to make them thermodynamically consistent.  Subsequently,
following the work of van Kampen [\rcite {vK}], Kac, Uhlenbeck and Hemmer
[\rcite{KUH}], Lebowitz and Penrose [\rcite {LP}] were able to derive
generalized 
mean field models in a more consistent mathematical way by considering
systems in which part of the interactions were explicitly modeled by Kac
potentials.  Kac potentials are simply potentials of the form $
\gamma^{d}U(\gamma x)$, which contain a range parameter $\gamma^{-1}$
[\rcite{KUH}], [\rcite{LP}].  For fixed $\gamma > 0$ these interactions are
essentially finite ranged but in the limit $\gamma
\to 0$ they become mean field like [\rcite{LP}].  Depending on whether
$\gamma^{-1}$ is small or comparable to the macroscopic scale (but always
very large compared to the interparticle distances) one gets either the 
usual MFT
(including the supplemental rules) or a macroscopic continuum theory (MCT)
from which one can also obtain the surface tension associated with phase
transitions caused by these Kac potentials [\rcite {GP}], [\rcite{ABCP}].  More 
recently it has
been possible to prove phase transitions close to the mean field ones for
large finite $\gamma^{-1}$ [\rcite{LMP}].

The structure of the coexisting phases in the MCT for a one component
system has been  extensively investigated in recent years by Presutti et al
[\rcite{DOPT}] and references therein. In multicomponent systems the MCT
is rather unexplored. In
the present work we study binary mixtures where the phase diagram has 
new qualitative
features, resulting from the possibility of segregation of the system into
regions rich in one of the species, e.g.\ oil and water.  Such systems were
already considered by van der Waals [\rcite{VW}] and by Korteweg 
[\rcite{K}] using MFT
expressions for the free energy. MFT however neglects the 
 spatial structures of the phase
domains so the starting point of our analysis is a MCT
expression for the free energy $\cal F$ of a binary mixture in a
macroscopic domain $\Lambda$, which we shall take for simplicity to be a
$d$-dimensional torus of side length $L$ and volume $|\Lambda| = L^d$.

The free energy functional that we consider satisfies certain
rearrangement inequalities,
and we study  the consequences of these for the phase diagram.
We will later focus on special cases of physical interest, such as a
mixture of van der Waals gases with a repulsive interaction
between the species. However, to make clear the role of the
rearrangements, we begin in a general setting in which these can be
applied.

Let $\rho_1$ and $\rho_2$ denote the densities for the two species, and
consider a ``free energy'' functional $\F$ that depends on $\rho_1$ and
$\rho_2$ through
$$\eqalign{\F(\rho_1,\rho_2) = \int_{\La}F(\r_1(x),\r_2(x)){\rm d}x &+
\int_{\La}\int_{\La}V(x-y)\rho_1(x)\rho_2(y){\rm
d}x{\rm d}y\ \cr &-
{1\over2}\int_{\La}\int_{\La}U(x-y)[\rho_1(x)\rho_1(y)+\rho_2(x)\rho_2(y)]{\rm 
d}x{\rm d}y\ .}\Eq(1.1)$$
The function $F$ represents the free energy density of a system with
only short range interactions. We require that $F$  be separately convex 
in $\r_1$ and $\r_2$, and that
the set of all $(\r_1,\r_2)$ for which $F$ is finite, which we call the
domain of $F$,  is an open convex set. Finally
we require that
$${\partial^2\over \partial \r_1\partial \r_2}F(\r_1,\r_2) \ge
0\Eq(cheddar1)$$
for all $(\r_1,\r_2)$ in the domain of $F$.
An example of particular interest is given by
$$F(\r_1,\r_2) = {1\over \b}\left[G(\r_1) + G(\r_2) +
D(\r_1+\r_2) -\mu_1\r_1 - \mu_2\r_2\right]\Eq(cheddar4)$$
where $\beta$ is the inverse temperature, $\mu_1$ and $\mu_2$ are 
chemical potentials,
 $G$ is a convex function, and $D$ takes account of short range
interactions between the particles which we have taken for simplicity to be
species independent.  Examples to keep in mind are:  
$$G(t) = t\log(t),\Eq(1.4a)$$
$$D(t)=\cases{-t\ \log (1-b t), \hbox{ if } x<b^{-1},\cr +\infty \hbox{
otherwise
}}\Eq(drho)$$
for some $b>0$, where $b^{-1}$ can be thought of as
the close packing density. In this case $D$ is defined on a bounded
interval, and the domain of $F$ is the set
$0 < \r_1,\r_2 < b^{-1}$. In particular the case $b=1$ corresponds to 
the lattice gas with
exclusion rule.

As for the long range interaction terms, we assume that  $V(x)$ and $U(x)$ are
non-negative radial decreasing functions of $x\in \R^d$ with compact
support and
$$
\int_{\R^d}V(x){\rm d}x = \alpha,\ \quad \int_{\R^d}U(x){\rm d}x = 
\sigma\ \Eq(al1def)$$

We further define
$$\ell = \int_{\IR}|x\big|[{1\over \alpha}V(x)+{1\over 
\sigma}U(x)\big]{\rm d}x  \
,\Eq(length)$$
which characterizes the length scale of the interactions.
We remark that it is not essential,  but
  convenient, to have the same attractive interaction $U$ for the two 
species.

Notice that $F$ is not required to be jointly convex or affine in $\r_1$
and $\r_2$, although this is expected on physical grounds, and, in any
case, in general $\F$ would not be because of the interaction
terms. Convexity of the interaction terms would require positive 
definiteness of the  $2\times 2$ matrix valued
function
$$\left(\matrix{-\hat U&\hat V\cr\hat V&-\hat U}\right),$$
where the hat
denotes the Fourier transform.. We
do not make such an assumption in this paper. Instead we make use in
our analysis of the fact, proven later, that the free energy functional
that we consider satisfies certain rearrangement inequalities. These give a
certain monotonicity to the spatial density profile of the two species.
We study the consequences of these for the phase diagram which originates
from the
competition between the repulsive
interaction terms, which would prefer to have the two
species  segregated as completely as possible, and the entropy term,
which would prefer to
have the densities of the two species be uniform over
$\Lambda$.

Our aim here is to study the minimizers of this functional either
under constraints on the integrals of $\ro$ and $\rt$,
or without such constraints.
Specifically, let $n_1$ and $n_2$ be two given positive numbers.  We 
define the sets ${\cal D}(n_1,n_2)$ and ${\cal D}$ of  constrained and 
unconstrained pairs of densities:
${\cal D}
(n_1,n_2)$ and ${\cal D}$ by
$${\cal D}(n_1,n_2) = \left\{(\rho_1,\rho_2)\in \R_{+}\times \R_{+}\ :\ 
{1\over
\vol}\int_{\La}\ro {\rm d} x = n_1\qquad{\rm and}
\qquad {1\over \vol}\int_{\La}\rt {\rm d}x = n_2\ \right\}\ ,\Eq(fracs)$$
$${\cal D}=\bigcup_{n_1,n_2\ge 0} {\cal D}(n_1,n_2).\Eq(fracs1)$$

The canonical (Helmholtz)  equilibrium free energy per unit volume or the
grand canonical (Gibbs)  equilibrium free energy per unit volume
  will be obtained by minimizing
${\cal F}(\rho_1,\rho_2)/|\La|$ under the constraint \equ(fracs),  or 
the unconstrained as in \equ(fracs1).  In the
former case the $\mu_i$ are irrelevant and we shall set them equal to zero.

The two cases will give different equilibrium density profiles $\bar
\rho_i(x)$, defined as the minimizer of $\cal F$ in (1.1), when there is a
coexistence of phases, i.e.\ when the minimizer for specified chemical 
potentials is not
unique.  This may happen when ${\cal F}(\rho_1,\rho_2)$ is not strictly
convex and the system undergoes a phase transition.

In the case where the diameter of the torus is large compared to $\ell$,
the interactions become approximately local. In this case,
for  densities $\rho_1$ and $\rho_2$ that are effectively constant on 
the length scale
$\ell$ over most of $\La$,  $\F(\rho_1,\rho_2)$
is well approximated by $\F_0(\rho_1,\rho_2)$ the MFT free energy
functional where
$$ \F_0(\rho_1,\rho_2) = \int_{\La}F(\r_1(x),\r_2(x)){\rm d}x + 
\alpha\int_{\La}\rho_1(x)\rho_2(x){\rm
d}x-{\sigma\over 2} \int_{\La}[\rho_1^2(x)+\rho_2^2(x)]{\rm d}x\ .
\Eq(delta)$$
When $\rho_1, \rho_2$ are restricted to be  constants, $n_1, n_2$ 
respectively, then ${\cal
F}_0(n_1,n_2)$ is essentially the function studied  by van der Waals [\rcite
{VW}] and by Korteweg [\rcite {K}]. 
(Korteweg however considered the
case $\a < 0$ so the rearrangement inequalities used here would not apply 
in his case.)
There are  interesting questions regarding the minimization of  $\F_0$ 
in  case
$\rho_1, \rho_2$ are not restricted to be  constant. 
For example, knowledge of the minimizing profiles 
enables one to compute
the equilibrium internal (interaction) energy at given temperature under 
the constraints in
\eqv(fracs). An application of this is given in Section 6.

We will show in section 2 that
$$\inf_{(\r_1,\r_2)\in {\cal D}}\F(\r_1,\r_2)\qquad{\rm and}\qquad
\inf_{(\r_1,\r_2)\in {\cal D}}\F_0(\r_1,\r_2)\Eq(infimum)$$
are both attained, and the same statement holds for the constrained problem.
Clearly
$$f_0 := {1\over |\La|}\inf_{(\r_1,\r_2)\in {\cal
D}}\F_0(\r_1,\r_2)\Eq(1.10)$$ is independent of $\La$. We shall see in
section 3 that, using of the rearrangement inequalities, we also have 
$$\lim_{|\La|\to \infty}{1\over |\La|}\inf_{(\r_1,\r_2)\in {\cal 
D}}\F(\r_1,\r_2)
= f_0\ ,\Eq(lim3)$$
and moreover there is a close relation between the minimizers in both 
infimums in \eqv(infimum).

The equality in \eqv(lim3) may look natural, but an example of Gates and 
Penrose  [\rcite{GP}] shows
that the  analogous statement  need not hold for simple one component 
systems. Specifically, their
one component system free energy functional is given by
$${\cal G}(\r) = {1\over \beta}\int_{\La}\r(x)\log \r(x){\rm d}x +
\int_{\La}\int_{\La}\rho(x)V(x-y)\rho(y){\rm
d}x{\rm d}y\Eq(1.12)$$
and
$${\cal G}_0(\r) = {1\over \beta}\int_{\La}\r(x)\log \r(x){\rm d}x +
\a\int_{\La}\int_{\La}\rho(x)^2{\rm
d}x\Eq(1.13)$$
where $\a$ and $V$ are related as before. Since $\a>0$, ${\cal G}_0$ is 
strictly convex, and hence is minimized at
the constant density $\r(x) = n$. However, if $\hat V(k_0)<0$ for some 
$k_0$, then for sufficiently small
$\e>0$, there is a $k$ close $k_0$ and a $\delta>0$ so that with 
$\r_\e(x) = n + \e\sin(kx)$,
$${1\over |\La|}{\cal G}(\r_\e) \le
{1\over |\La|}{\cal G}_0(n)- \delta \Eq(1.14)$$
for all $\La$ sufficiently large.
In our model, the rearrangement inequalities prevent oscillations from 
lowering the free energy.

In more detail, observe that
$$
{\cal F}(\rho_1,\rho_2) = {\cal F}_0(\rho_1,\rho_2) - {1 \over 2}
\sum_{i,j} \int \int_\Lambda \xi_i (x,y) V_{ij}(x-y) \xi_j(x,y) {\rm d}x 
{\rm d}y
\Eq(differ)
$$
where
$$
V_{11} = V_{22} = -U\qquad{\rm and}\qquad V_{12} = V_{21} = V\ ,\Eq(1.16)$$
and
$$\xi_i(x,y) = \rho_i(x) - \rho_i(y)\ .\Eq(1.17)$$
The rearrangement inequalities will ensure that  final integral in  the 
right of \equ(differ) is a surface term.
This will be proved in Section 3 below.

The rearrangement inequalities have other interesting
consequences for the phase diagram of our system. To be concrete, we will
investigate these in the case of the van der Waals free energy 
functional given by \eqv(cheddar4) and \eqv(drho).
Because of \eqv(lim3), the rearrangement inequalities allow us to draw 
conclusions about the phase diagram
of $\F_0$ even though  rearrangement of $\r_1$ and $\r_2$ may not affect 
the value of
$\F_0(\r_1,\r_2)$.

We will see from the structure of the minimizers that the system,
for appropriate choices of $F$, $V$ and $U$,
exhibits two phase transitions: A {\it liquid-vapor} transition
and a {\it segregation} transition. In the first of these, below
a precisely determined critical temperature, the system separates
into two regions, one a {\it liquid region} in which the total bulk density
is uniformly
very close to some number $\rho_\ell$, and the other a {\it vapor
region} in which the total density is
very close to some number $\rho_v$. This is similar to a van der Waals
type transition in a one component system.  
The other transition is a segregation transition.  The liquid or vapor 
homogeneous phases can
split into two sub regions
in which the densities  of the
two species are very close to certain preferred concentrations 
$\rho_\ell^+$ and $\rho_\ell^-$ for the liquid,
and $\rho_v^+$ and $\rho_v^-$ for the vapor.
The values of these densities depend on
the temperature and the parameters of the system, and a fairly detailed 
analysis is required to determine them in any given case.
Nonetheless, there are general conclusions we can draw as a consequence 
of the rearrangement inequalities;
for example, we will see in broad generality that for our system one
always has
$$\r_\ell^- \le \r_v^- \le \r_v^+ \le \r_\ell^+\Eq(1.18)$$
and never
$$\r_v^- \le \r_v^+ \le \r_\ell^- \le \r_\ell^+\ .\Eq(1.19)$$

The paper is organized as follows: In section 2, we state precisely the
rearrangement inequalities referred to above, and apply them to prove the
existence of minimizers.  We also prove that minimizers necessarily have
certain regularity properties. In section 3 we prove a general theorem
showing that as a consequence of the rearrangement inequalities, minimizing
$\F$ and minimizing $\F_0$ are essential by the same problem for $L$ much
larger than $\ell$. The general theorem requires certain assumptions on
$F$.  These assumptions are verified for the physical model that we study
in the next two sections. Section 4 is devoted to the application of convex
analysis methods to deduce general features of the minimizers for our
system. We will see that there are generically one, two, three or four
coexisting equilibrium states in this model. The general convex analysis
does however permit more bizarre coexistence of phases. In section 5 we
impose further restrictions (physically natural) and obtain a detailed
phase portrait, ruling out these bizarre non--generic behaviors.  Section 6
gives an application of the results obtained here to the study of the
equilibrium states of an energy conserving kinetic model.  In particular,
the results obtained here allow us to determine the equilibrium partition
of the total energy into kinetic and interaction terms.  Thus, given the
energy, we are able to determine the equilibrium temperature in such a
system.

\bigskip
\bigskip
\chap {2. Existence and Regularity of Minimizers.} 2
\numsec= 2
\numfor= 1
\numtheo=1
\medskip

In this section we  show that minimizing densities exist, and  begin the
determination of their nature.
Since particles of the same species
attract and of different species repel, one can expect that the minimizing
densities should be
``well organized'' in some sense, and we  use a rearrangement inequality
to prove this.

Introduce coordinates
$x_i$, $i = 1,\dots, d$ with
$-L \le x_i < L$ for each $i = 1,\dots, d$ where  $L > 0$, and
we make the usual
identifications to parametrize our torus $\La$.

We say that a function $g$ on $\La$ is {\it symmetric
monotone decreasing} about the origin $0,\dots, 0$ if it is a symmetric
decreasing function of each coordinate
$x_i$ for  $-L \le x_i < L$. That is, for each $i$
$$g(x_1,x_2,\dots ,x_i,\dots x_d) = g(x_1,x_2,\dots , -x_i,\dots
x_d)\Eq(2.1)$$
and whenever $L \ge y_i>x_i\ge 0$, then
$$g(x_1,x_2,\dots ,y_i,\dots x_d) \le  g(x_1,x_2,\dots , x_i,\dots x_d)\
.\Eq(2.2)$$
We define {\it symmetric
monotone increasing} in the analogous way. Finally we say that a 
function is symmetric monotone if it is either
symmetric monotone decreasing or increasing.

Given a non--negative integrable function $g$  on $\Lambda$, we say that 
  a  non--negative integrable
function $ g^\star$ is a symmetric monotone decreasing rearrangement  of 
$ g$ if it is a symmetric monotone
decreasing function such that  for all $t>0$
$\{x | g(x) > t\}$ has the same measure as $\{x | g^\star(x) > t\}$. 
Analogously we define a symmetric monotone
increasing rearrangement $h_\star$ of a non--negative integrable
function $h$.

We will see below that for an appropriate choice of $\rho_1^\star$ and 
$\rho_{2\star}$,
$\F(\rho_1^\star,\rho_{2\star})\le \F(\rho_1,\rho_{2})$, with equality 
only in case $\rho_1$ and $\rho_2$ are
already symmetric monotone up to a common translation.
  The relevance of these definitions lies in the following lemma, which 
says that the
``total variation'', $\|\nabla g\|_1$,
of a symmetric monotone function $g$ is a ``surface term''. (Here and in
what follows, $\|\cdot\|_p$ denotes the $L^p(\La)$
norm.)
\medskip
\noindent{\bf Lemma 2.1} {\it Suppose that $g$ is a bounded and
non-negative integrable
function on $\La$ that is symmetric monotone. Then  $g$ has an
integrable distributional gradient,
and the total variation of $g$, that is $\|\nabla g\|_1$, satisfies
$$\int_{\La}\left|\nabla g(x)\right|{\rm d} x \le
\vol^{(d-1)/d}2d\|g\|_\infty\ .
\Eq(totv)$$ }
\medskip
Note that the right hand side is proportional to $\vol^{(d-1)/d}$, which
is what makes it a surface term.
\medskip
\noindent{\bf Proof:} We suppose first that $g$ is smooth and symmetric 
monotone decreasing, and derive
the bound \eqv(totv)
under this hypothesis.
  Any  such  function $g$  has the
   property that as one goes around any of the circles where only one
coordinate varies, say $x_i$,
one proceeds monotonically from the minimum of $g$ on this circle, to
the maximum, and then
back down to the minimum. Since $g\ge 0$, the difference between the
maximum and the
minimum is no more than $\|g\|_\infty$. Hence, since $g$ is smooth,
$$\int_{\La}\left|{\partial g\over \partial x_i}\right|{\rm d} x \le
{\vol\over L}
\left(2\|g\|_\infty\right)\ .\Eq(2.4)$$
Now integrating over the remaining
coordinates one obtains \eqv(totv).

For the general case, one may approximate $g$ by applying the heat
kernel on the torus to it.
This approximation preserves both the positivity and the property of
being symmetric monotone.
Also, the heat evolution is continuous in both the total variation norm
and the $L^\infty$ norm,
so \eqv(totv) is preserved as the approximation is relaxed. \eop
\medskip

This lemma will play a
crucial role in determining the structure of our minimizers, since,
as we will see below, our minimizers are symmetric monotone, up to
translation on the
torus, and thus, they satisfy the bound \eqv(totv) as well as {\it
a--priori} $L^\infty$
bounds.

\medskip
\noindent{\bf Lemma 2.2} {\it Let $F$ be a  function on
$\R_+\times \R_+$ that
satisfies \eqv(cheddar1) everywhere on its domain. Let $U$ be a
non-negative strictly monotone decreasing radial function on
$\R^d$. Suppose also that the diameter of its support is less than the
smallest period of $\La$. Then
for all non--negative functions $g$ and $h$ on $\Lambda$, there is a
symmetric monotone decreasing
function $g^\star$ and a
symmetric monotone increasing $h_\star$ so that
$$\int\int_{\La}g(x)U(x-y)h(y){\rm d}x{\rm d}y \ge
\int\int_{\La}g^\star(x)U(x-y)h_\star(y){\rm d}x{\rm d}y \Eq(reran1)$$
and
$$\int_{\La}F(g(x),h(x)) {\rm d}x \ge
\int_{\La}F(g^\star(x),h_\star(x)) {\rm d}x\ .\Eq(reran2)
$$
Moreover, there is equality in \eqv(reran1)  only in
case $g=g^\star$ and $h=h_\star$ up to a common translation.}
\medskip

These estimates \eqv(reran1) could be deduced from related rearrangement
inequalities of Baernstein and
Taylor [\rcite{BT}], and  Luttinger [\rcite{L}]. However,  a direct 
proof is provided in Appendix A.
The second inequality is reminiscent of an
inequality of Almgren and Lieb [\rcite{AL}] with the opposite sense for 
symmetric
decreasing rearrangement on $\R^n$ in which both functions are
``piled up'' around the {\it same} point. The proofs though are
different.

\medskip

We now turn to the existence of minimizers for the variational problem
of determining
$$\inf\left\{ {\cal F}(\rho_1,\rho_2)\ :\ (\rho_1,\rho_2)\in {\cal
D}(n_1,n_2)\ \right\}\Eq(2.6)$$
where ${\cal D}(n_1,n_2)$ is given by \eqv(fracs).

The lemmas collected above enable this to be proved in a wide variety of
circumstances. Theorem 2.3 below
applies to one of these that is of particular physical interest.
Mathematically, it is a very direct consequence of the
lemmas since when $D$ is given by \eqv(drho), there is a trivial {\it
a-priori} bound on $\|\rho_1\|_\infty$
and  $\|\rho_2\|_\infty$ -- namely $b^{-1}$.

Without such a term to enforce {\it a--priori} $L^\infty$ bounds, and
without an entropy that
prevents a vacuum, the proof is more involved, but still works.
Mathematically, this context is much more interesting,
if less physical. To keep the focus on the main physical case, we prove 
here
the theorem only in the simple case required to treat
the van der Waals gas and put in Appendix B the statement and the proof 
for the other case.

The role of the rearrangement inequalities is this: Since the interaction
is not positive definite, lowering the
interaction energy  may well favor
oscillations in the densities $\r_1$ and $\r_2$, and there is no
mechanism to prevent a minimizing sequence from
oscillating more and more rapidly so that some part of the mass vanishes
in a weak limit, with the consequence that
the limit no longer belongs to ${\cal D}(n_1,n_2)$. This cannot happen
here because of Lemma 2.2.

\bigskip
\noindent{\bf Theorem 2.3} {\it Let $F$ be given by \eqv(cheddar4),
where in particular
$D$ has the property that $D(t) = \infty$ for $t > b^{-1}$ for some $b>0$.
Then, for all $n_1$ and $n_2$ and all $\beta$,
$$\F(n_1,n_2,\beta) = \inf_{(\ro,\rt)\in {\cal
D}(n_1,n_2)}\F(\ro,\rt)\Eq(minA)$$
is achieved at least at one minimizing pair $(\ro,\rt)$. Any such
pair satisfies the
Euler--Lagrange equations
$$G'(\ro) + D'(\ro+\rt) +\beta V*\rt -  \b U*\rho_1 =
C_1\quad {\rm and}\quad G'(\rt) + D'(\ro+\rt) +\beta V*\ro -
\b U*\rho_2 = C_1\ .\Eq(elA) $$
Moreover,
$$\|\nabla \ro\|_1 \ ,
\|\nabla \rt\|_1\ \le |\Lambda|^{(d-1)/d}2d b^{-1}
\ ,\Eq(linf3A)$$
and if $(\ro,\rt)$ is any minimizing pair, then
$\ro=\ro^\star$ and $\rt={\rt}_\star$ up to a common translation.

Likewise, for any $\mu_1$, $\mu_2$, let
$$ \F(\mu_1,\mu_2,\beta)= \inf_{(\ro,\rt)\in {\cal
D}}\left\{\F(\ro,\rt)-\mu_1\int_\La \ro(x){\rm d}x -\mu_2\int_\La 
\rt(x){\rm d}x\right\}\
.\Eq(2.11)$$
Then the minimizers do exist and satisfy the above conditions with $C_1$ and
$C_2$ in \equ(elA) replaced by
$\mu_1$, $\mu_2$.}
\bigskip
\noindent{\bf Proof:} We first observe that we may assume $\ro,\rt \le
b^{-1}$.
These {\it a--priori} bound makes this case easier.

The
functional
$$\ro,\rt \mapsto  \F(\ro,\rt)\Eq(GGdefA)$$
is jointly continuous in the $L^1$ topology by the dominated convergence
theorem.

Now let $(\ro^{(k)},\rt^{k)})$ be a minimizing sequence satisfying  the
constraint \eqv(fracs)$_1$, and such that
$(\ro^{(k)},\rt^{k)})$ is a symmetric monotone pair. Such a sequence
exists by  Lemma 2.2.
By the Helly selection principle and the {\it a--priori} pointwise
bound, this sequence is strongly compact in
$L^1\times L^1$. Passing to a convergent subsequence, we have our
minimizing pair.

The Euler--Lagrange equations now easily follow, and then \eqv(linf3A)
follows from \eqv(totv) and the {\it a--priori}
sup norm bound provided by $D$.\quad \eop
\medskip
\medskip

\chap {3. Large Volume.} 3
\numsec= 3
\numfor= 1
\numtheo=1

In this section we focus on the case in which $L$, the period of the 
torus $\Lambda$, is large compared to
$\ell$, the length scale of the interaction defined in \equ(length).

\bigskip
\noindent{\bf Lemma 3.1} {\it Let $(\ro,\rt)$ be any minimizer for
the free energy functional. Then, using $*$ for convolution,
$$\sum_{j=1}^2 \int_{\La}\left|V*\r_j -
\alpha \r_j\right|{\rm d}x + \sum_{j=1}^2 \int_{\La}\left|U*\r_j -
\s \r_j\right|{\rm d}x
\le C\vol^{(d-1)/d}\Eq(aflat)$$
for some constant $C$ depending only on $\beta$ and $n_1,n_2$. As a
consequence,
$$\sum_{j=1}^2\left|\int_{\La} \r_i V*\r_j {\rm d} x -
\a \int_{\La} \r_i \r_j
{\rm d} x\right| +\sum_{j=1}^2\left|\int_{\La} \r_i U*\r_j {\rm d} x -
\s \int_{\La} \r_i \r_j
{\rm d} x\right|
\le C\vol^{(d-1)/d}\Eq(almost)$$
where again, $C$ depends only on  $\beta$ and $n_1,n_2$.}
\bigskip
\noindent{\bf Proof:}
$$\r_j*V(x) -\a\r_j(x) = \int_\La
V(y)\left[\r_j(x-y) -
\r_j(x)\right]{\rm d}y =
-\int_0^1 \int_\La V(y)\nabla \r_j(x-ty)\cdot y{\rm d}y{\rm d}t\ .\Eq(3.3)$$
Hence
$$\int_\La \left|\r_j*V(x) -\a\r_j(x)\right|{\rm d}x \le
\left(\int_{\La}V(y)|y| {\rm d}y\right)\|\nabla \ro\|_1\ ,\Eq(3.4)$$
and from Lemma 2.1,
$$\int_\La \left|\r_j*V(x) -\a\r_j(x)\right|{\rm d}x \le
C\left(\int_{\La}V(y)|y|{\rm d}y\right)\vol^{(d-1)/d}\ .\Eq(3.5)$$
The same argument applies to the terms involving $U$.
\eop
\bigskip

Now fix any $\epsilon>0$. A simple Chebyshev argument based on the
integral bounds of
Lemma 3.1 shows that, by choosing $\L$ sufficiently large,  off of a set
of ``surface term size'', $V*\r_i =
\a\r_i \pm
\epsilon$ and
$U*\r_i = \s\r_i \pm \epsilon$. More precisely, with $|A|$ denoting
the Lebesgue measure of a set $A$,
$$\left|\{x\ |\ |V*\r_i(x) - \alpha\r_i(x)| \ge \epsilon\
\}\right| \le C\e^{-1}
\ell\vol^{(d-1)/d}, \quad
i=1,2\ ,\Eq(3.6)$$
and likewise for $U$.

Let $G_\epsilon$ be the ``good'' set
$$\left(\bigcap_{j=1}^2 \{x\ |\ |V*\r_j(x) - \a\r_j(x)| <
\epsilon\ \}\right)\bigcap
\left(\bigcap_{j=1}^2 \{x\ |\ |U*\r_j(x) - \s\r_j(x)| <
\epsilon\ \}\right)
\ .\Eq(gsetdef)$$

Any minimizing pair of densities $(\r_1,\r_2)$ for $\F$ with $n_1$ and 
$n_2$ fixed
must satisfy the Euler Lagrange equations
$$\eqalign{
&{\partial F\over\partial  \r_1}(\ro,\rt) + V*\rt-  U*\ro=C_1\cr
&{\partial F\over \partial \r_2}(\ro,\rt)+ V*\ro- U*\rt=C_2}\Eq(cafe)$$
for some $C_1$ and $C_2$.

Likewise, any minimizing pair of densities $(\tilde\r_1,\tilde\r_2)$ for 
$\F_0$ with $n_1$ and $n_2$ fixed
must satisfy the Euler Lagrange equations
$$\eqalign{& \tilde C_1={\partial F\over\partial 
\tilde\r_1}(\tilde\ro,\tilde\rt)+ \a\tilde\rt- \s\tilde\ro, \cr&
  \tilde C_2={\partial F\over\partial  \tilde\r_2}(\tilde\ro,\tilde\rt)+ 
\a\tilde\ro- \s\tilde\rt
\ .}\Eq(elltrue2) $$

Given a minimizing pair $(\r_1,\r_2)$ for $\F$,
define functions $C_1(x)$ and $C_2(x)$ through
$$\eqalign{
&C_1(x):={\partial F\over\partial  \r_1}(\ro,\rt)+ \a\rt- \s\ro, \cr&
C_2(x):={\partial F\over\partial  \r_2}(\ro,\rt)+ \a\ro- \s\rt
\ .}\Eq(elltrue) $$
It follows from the definitions that on $G_\epsilon$
$$|C_1(x) - C_1|\le \epsilon\qquad{\rm and}\qquad
|C_2(x) - C_2|\le \epsilon\ .\Eq(close)$$
Therefore,  on the set $G_\epsilon$, $(\ro(x),\rt(x))$ is a
solution of
\equ(elltrue)
with $C_1(x)$ and $C_2(x)$ very close to $C_1$ and $C_2$. We wish to 
conclude that
  on $G_\epsilon$, the values of $(\ro(x),\rt(x))$ are essentially
those of a minimizer of $F_0$.

Toward this end, we  make a ``stability'' assumption on the
Euler Lagrange equations  describing
minimizers of $\F$ under the constraint for fixed $n_1$ and $n_2$. This 
condition is easy
to verify for many particular choices of $F$, as we shall see.
\bigskip
\noindent{\bf Definition: (Amenable Free Energy Function)}
We say that the free energy function
$F(\r_1,\r_2) + \a\r_1\r_2 - (\r_1^2+\r_2^2)$ is {\it amenable} when
for  any given
constants $ C_1$ and
$ C_2$, there are at most a finite number $N$ of  pairs of numbers
$$(\r_1^{(i)},\r_2^{(i)})\qquad i=1,\dots, N\Eq(3.12)$$
so that
for any given
$\epsilon>0$, there is a $\delta>0$ depending only on $\epsilon$, $C_1$ 
and $C_2$
such that the following is true:

Whenever any pair of numbers $(\tilde \r_1,\tilde \r_2)$ satisfies
$${\partial F\over\partial  \r_1}(\tilde \ro,\tilde\rt)+ \a\tilde\rt- 
\s\tilde\ro =
\tilde C_1\qquad {\rm and}\qquad
{\partial F\over\partial  \r_2}(\tilde \ro,\tilde \rt)+
\a\tilde\ro-\s\tilde\rt =\tilde C_2\ .\Eq(mango5)$$
for some $\tilde C_1$ and $\tilde C_2$ with
$|\tilde C_1 - C_1| + |\tilde C_2 - C_2| < \delta$,
it follows that
$$|\tilde \r_1-\r_1^{(i)}|+|\tilde \r_2-\r_2^{(i)}|<\e, \quad {\rm for\ 
some}\quad 1 \le i \le N\
.\Eq(3.13.5)$$
\bigskip
Checking this in practice amounts to checking  that there is continuous 
dependence
of  solutions to \equ(elltrue2) from $(C_1,C_2)$. Just to concrete, 
consider the simple case in which
$F(\r_1,\r_2) = (\r_1\log \r_1 +  \r_2\log \r_2)/\b$. Then 
\eqv(elltrue2) are
$$\log \ro +\b\a\rt = C_1\qquad {\rm and}\qquad \log \rt +\b\a\ro = 
C_2\Eq(el54) $$
Then with
$$h_{C_1,C_2}(\r) = e^{C_1}\exp(-\b\a e^{C_2}(\exp(-\b\a\r)))\ ,$$
$\r_1$ and $\r_2$ satisfy the fixed point equations
$$\r = h_{C_1,C_2}(\r)\qquad{\rm and}\qquad \r = h_{C_2,C_1}(\r)$$
respectively. For each $C_1$ and $C_2$ there is a number $R$  depending 
only on $C_1$ and $C_2$
so that $h_{C_1,C_2}(\r)$ is convex for $\r < R$, and concave for 
$\r>R$. One sees that
there are always one, two or three solutions to \eqv(el54) in this case. 
As $C_1$ and $C_2$ vary
in small intervals, there are at most three small pairs intervals 
required to hold all of the solutions.
This is the situation described in general by the definition.

We will give further physical examples in the next section, and here
we focus on the general consequences of amenability.

The main consequence is that on the ``good set''  $G_\epsilon$, any 
minimizers $\ro$ and $\rt$ of $\F$
are essentially
discrete, taking
their values in the union of a finite number of short intervals. We now 
identify these intervals, and relate
the minimization of $\F$ and $\F_0$.
\bigskip
\noindent{\bf Theorem 3.2} {\it  Suppose that the free energy  function 
is amenable.
For any fixed $(n_1,n_2)$,
let $(\r_1,\r_2)$ be a minimizer for
${\cal F}$ in ${\cal D}(n_1,n_2)$. Then there is a finite set of pairs 
of numbers
$$( \r_1^{(i)}, \r_2^{(i)})\qquad i=1,\dots, N\Eq(3.14)$$
such that for all
$\epsilon>0$, and all $L$ sufficiently large,
there is
a set $G\subset \L$
such that
$${|G|\over |\L|}< C|\L|^{-1/d}$$ such that
for all $x$ in $G_\epsilon$,
$$|\r_1(x)- \r_1^{(i)}| + |\r_2(x)- \r_2^{(i)}|<\e\Eq(aba)$$
for some $1\le i \le N$.
Moreover, there is a pair $(\tilde \r_1,\tilde \r_2)$ satisfying
$|\tilde\r_1- \r_1^{(i)}| + |\tilde\r_2- \r_2^{(i)}|<\e$ that satisfies the
Euler--Lagrange equation \eqv(elltrue2) for a
minimizer of $\F_0$ in ${\cal D}(\tilde n_1,\tilde n_2)$ for some 
$\tilde n_1$ and
$\tilde n_2$ with $|\tilde n_1- n_1| + |\tilde n_2- n_2| < \e$.
}
\bigskip
\noindent{\bf Proof:}
For some $\k$ to be determined later, let $G = G_\k$.
Then,  as a consequence of Lemma 3.1, $\r_1(x)$ and $\r_2(x)$
satisfy \eqv(elltrue) for $C_1(x)$ and $C_2(x)$ nearly
constant on $G$. Let $C_1$ and $C_2$ be the respective average values of 
$C_1(x)$ and $C_2(x)$ on $\La$.
It follows that on $G$, $\r_1$ and $\r_2$ have values lying finite 
number of pairs of intervals.
As $\kappa$ is decreased, so is the width of these intervals. Each pair 
of values in these intervals is
a solution of the Euler--Lagrange equations \eqv(elltrue2) for some 
values of $\tilde C_1$ and
$\tilde C_2$ close to $C_1$ and $C_2$ respectively. By decreasing 
$\kappa$, we can ensure that
$|\tilde C_1-C_1| +|\tilde C_2-C_2| < \delta $, where $\delta$ is 
related to $\e$ as in the definition of
amenability.
\eop

We remark that in specific cases, it is possible to carry the analysis 
further, and to show that
the pairs of values $(\tilde \r_1,\tilde \r_2)$ correspond to values of 
minimizers for $\F_0$,
and not simply solutions of the Euler--Lagrange equation. At this level 
of generality, that is not possible.

In this section we have only considered the constrained problem because 
in the unconstrained case the situation
is much simpler. In the unconstrained case surface tension would 
discourage partitioning $\Lambda$ among
different minimizing phases.

\bigskip
\medskip
\medskip
\chap {4. Local Interactions.} 4
\numsec= 4
\numfor= 1
\numtheo=1

\bigskip

In this section we consider the case in which  $V(x)=\alpha\delta(x)$ 
and  $U(x)=\sigma\delta(x)$.
This will enable us to obtain a very complete picture of the
minimizers when $\s$ is sufficiently small and a rather clear
understanding for arbitrary $\s$,
as we show in the next section. Note that this corresponds exactly to 
replacing
$\cal F$ by ${\cal F}_0$.

We begin our analysis by looking for the unconstrained spatially
homogeneous
minimizers of ${\cal F}_0$  with additional chemical potentials $\mu_1$
and $\mu_2$
({\it grand-canonical ensemble}) which will play the role of Lagrange
multipliers later
on. Therefore, fixing $\mu_1$ and $\mu_2$, we look at the minimizers of
grand canonical
free energy density on
$\R_+\times \R_+$ given by
$$f_{\mu_1,\mu_2}(\ro,\rt)=
F(\ro,\rt)+\a\ro\rt-{\s\over
2}(\ro^2+\rt^2)-\mu_1\ro-\mu_2\rt
\Eq(function)$$

Since this model exhibits both condensation-evaporation transition and 
segregation transition the natural variables are not
$ \rho_1$ and $ \rho_2$ but $ \rho$ and $\phi$ where
$$\rho(x) = \rho_1(x)+\rho_2(x)\qquad{\rm and}\qquad \phi(x) =
{\rho_1(x) - \rho_2(x)\over \r(x)}\ .\Eq(4.2)$$
Then the free energy ${\cal F}_0(\rho_1,\rho_2)$ can be written as
$${\cal F}_0(\rho_1,\rho_2) = \int_\Lambda
g_{\mu,h}(\rho(x),\phi(x)){\rm d}x\ ,\Eq(4.3)$$
where
$$g_{\mu,h}(\rho,\phi) = F\left({\rho\over 2}(1+\phi),{\rho\over
2}(1-\phi)\right)
+{\a\over 4}\rho^2(1-\phi^2)- {\s\over 4}\rho^2(1+\phi^2)
-\mu\rho-h\rho\phi
\Eq(seg55)$$
where
$$\mu={\mu_1+\mu_2\over 2}, \quad h={\mu_1-\mu_2\over 2}.\Eq(4.5)$$

It turns out that, under our assumptions, minimizers of this functional
have a very special structure: Each of $\rho_1$
and $\rho_2$ can take on at most four values. The region $\Lambda$ is
decomposed into at most four subregions
in which both $\rho_1$ and $\rho_2$ are constant. The four possible
values of the densities
result from a possible condensation-evaporation transition, together 
with  a
possible segregation transition.

Let $\tilde g_{\mu,h}$ be the {\it convex minorant} of
$g_{\mu,h}$. That
is,
$$\tilde g_{\mu,h}(\rho,\phi)=\sup_{\ell\in {\cal L}} \{\ell(\rho,\phi)\ 
| \ \ell(\rho,\phi)\le
g_{\mu,h}(\rho,\phi)\}\Eq(4.6)$$
where ${\cal L}$ is the set of all the linear functions
$\ell(\rho,\phi)=a\rho + b\phi+c,\ a,b,c \in \R$.
Let $B$ be the set of $\rho$ and $\phi$ for which
$$g_{\mu,h}(\rho,\phi) > \tilde g_{\mu,h}(\rho,\phi)\ .\Eq(4.7)$$
$B$ is the set where there are ``flat spots'' in the graph of
$\tilde g_{\mu,h}$, and as is clear and well known, these are
irrelevant to
the minimization problem so that
$$\inf_{\rho_1,\rho_2}{{\cal F}_0(\rho_1,\rho_2)\over |\Lambda|} =
\inf_{\rho,\phi}{1\over
|\Lambda|}\int_\Lambda\tilde g_{\mu,h}(\rho(x),\phi(x)){\rm d}x\ .\Eq(4.8)$$

Next, define
$$\psi_{\mu,h}(\rho)=\inf_{\phi\in [-1,1]}
g_{\mu,h}(\rho,\phi)\ ,\Eq(hat)$$
and
$$\tilde \psi_{\mu,h}(\rho)=\inf_{\phi\in
[-1,1]}\tilde g_{\mu,h}(\rho,\phi). \Eq(hat1)$$
Then  $\tilde \psi_{\mu,h}$ is the convex minorant of $
\psi_{\mu,h}$. Indeed, the epigraph of $\tilde g_{\mu,h}$;
i.e., the set
$$\{(\rho,\phi,z)\ :\ z \ge \tilde g_{\mu,h}(\rho,\phi)\ \}\Eq(4.11)$$ is
convex, and the  epigraph of $\tilde \psi_{\mu,h}$ is simply
the projection
of the epigraph of $\tilde g_{\mu,h}$ onto the $\rho,z$ plane.

It is clear that
$${1\over |\Lambda|}\int_\Lambda
\tilde g_{\mu,h}(\rho(x),\phi(x)){\rm d}x \ge
{1\over |\Lambda|}\int_\Lambda
\tilde \psi_{\mu,h}(\rho(x)){\rm d}x\Eq(4.12)$$
with equality if and only if for almost every $x$,
$\phi(x)$ minimizes $\tilde g_{\mu,h}(\rho(x),\phi)$ considered as
a function of
$\phi$.

Now let $\rho_{\rm ave}$ be given by
$$\rho_{\rm ave} = {1\over |\Lambda|}\int_\Lambda \rho(x){\rm d}x\ 
.\Eq(4.13)$$
Then by Jensen's inequality,
$${1\over |\Lambda|}\int_\Lambda
\tilde \psi_{\mu,h}(\rho(x)){\rm d}x \ge
\tilde \psi_{\mu,h}(\rho_{\rm ave})\ .\Eq(4.14)$$
It is well known that if $\tilde \psi_{\beta,\mu,h}$ is strictly convex,
then there is
equality if and only if $\rho(x) = \rho_{\rm ave}$ almost everywhere.
However, even if
$\tilde \psi_{\mu,h}$ is not strictly convex, the proof of
Jensen's inequality
has important consequences for our problem.

To explain these, we first recall that if $\psi$ is any convex function
on the
positive axis, then an affine function $\ell(\rho)= a\rho + b$ is a {\it
supporting
line} for $\psi$ in case $\psi(\rho) \ge \ell(\rho)$ for all $\rho$, and if
$\psi(\rho_0) = \ell(\rho_0)$ for some $\rho_0$. If $\ell$ is any
supporting line for
$\psi$, then the set
$$\{\rho \ :\ \ell(\rho) = \psi(\rho)\ \}\Eq(4.15)$$
is a closed interval. Such an interval is called a {\it support
interval} of $\psi$.
The following is simply Jensen's inequality,  with the only novel
feature being that
the  statement about the cases of equality that applies outside the
strictly convex case.
\medskip
\noindent{\bf Lemma 4.1} {\it Let $(\Omega,{\cal S},\nu)$ be a
probability measure space, $\psi$
a convex function on the positive axis, and $\rho$ a non--negative
measurable function.
Then
$$\int \psi(\rho){\rm d}\nu \ge \psi\left(\int \rho{\rm d}\nu\right)\ 
,\Eq(4.16)$$
and there is equality if and only if, up to a set of measure zero,
$\rho$ takes it values in a
single support interval of $\psi$.}
\medskip

\noindent{\bf Proof:} This follows from a close examination of the
standard proof of Jensen's inequality, which turns on
the fact that if $f$ and $g$ are two measurable functions,
$$\int \max\{f,g\}{\rm d}\nu \ge \max\left\{\int f {\rm d}\nu\ ,\ \int g
{\rm d}\nu\ \right\}\ ,\Eq(4.17)$$
with equality if and only if either $f = \max\{f,g\}$ or $g =
\max\{f,g\}$ almost everywhere.
This applied to
$$\psi(\rho) = \sup_{\ell}\ell(\rho)\Eq(4.18)$$
where the supremum is taken over all supporting lines of $\psi$ yields
the inequality. By the
above, there is equality if and only if for any two supporting lines
$\ell_1$ and $\ell_2$,
either $\ell_1(\rho(x)) = \max\{\ell_1(\rho(x)),\ell_2(\rho(x))\}$ or
$\ell_2(\rho(x)) = \max\{\ell_1(\rho(x)),\ell_2(\rho(x))\}$ for almost
every $x$. This can only
happen if, almost everywhere, $\rho$ takes on all of its values in a
single supporting interval.\quad \eop

This has the immediate consequence that for any minimizer of the local
interaction problem, the values of $\rho(x)$
all lie in a single support interval of $\tilde \psi_{\mu,h}$. Now
in most cases that we will
consider,
$$\psi_{\mu,h}(\rho) > \tilde\psi_{\mu,h}(\rho)\Eq(env)$$
for all $\rho$ in any support interval, except at the endpoints. In
particular, this is the case if
$\psi_{\mu,h}$ is almost everywhere either strictly convex or
strictly concave, and there are no lines that are tangent
at three or more points.  Then a minimizer for our problem cannot have
$\rho(x)$ in the interior of the support
interval because of \eqv(env). In exceptional cases, there may be
physical values in the interior of a support interval.
This occurs when there is a point of triple tangency, or higher.

Putting aside this exceptional case for the moment,  there are only two
possibilities for any minimizer in our problem: Either
\medskip
\noindent{$\bullet$}  {\it The support interval of $\tilde
\psi_{\mu,h}$ that contains
$\rho_{\rm ave}$ consists of $\rho_{\rm ave}$ alone. In this case,
$\rho(x) = \rho_{\rm ave}$
almost everywhere.}
\medskip
\noindent{$\bullet$}  {\it The support interval of $\tilde
\psi_{\mu,h}$ that contains
$\rho_{\rm ave}$ consists of a closed interval $[\rho_v,\rho_\ell]$ with
$\rho_v < \rho_\ell$.
In this case, $\Lambda = \Lambda_v \cup \Lambda_\ell$ with
$\rho(x) = \rho_v$ almost everywhere in $\Lambda_v$ and
$\rho(x) = \rho_\ell$ almost everywhere in $\Lambda_\ell$.}
\medskip

We say that there is a {\it condensation-evaporation} transition in the 
second
case; the region $\Lambda_v$
contains the vapor state and $\Lambda_\ell$ contains the liquid state.
The volumes of these two regions
are given by
$$|\Lambda_\ell|\rho_\ell + |\Lambda_v|\rho_v = |\Lambda|\rho_{\rm ave}
= n_1+n_2\ .\Eq(volU)$$
Note that the volume fractions of the liquid and vapor states are
determined by
this relation alone, before we begin considering any possible
segregation in either the
vapor or the liquid.

We note that there is a condensation-evaporation transition exactly where
$\psi_{\mu,h}$ has an interval of
concavity. Any such interval is contained in an interval
$[\rho_v,\rho_\ell]$ with
$$\psi_{\mu,h}(\rho_v)  =
\tilde\psi_{\mu,h}(\rho_v)\qquad{\rm and}\qquad
\psi_{\mu,h}(\rho_\ell)  = \tilde\psi_{\mu,h}(\rho_\ell)\ \Eq(4.21)$$
and
$$\psi'_{\mu,h}(\rho_v) = \psi'_{\mu,h}(\rho_\ell)\ .$$
This means that
$$\int_{\rho_v}^{\rho_\ell}\psi''_{\mu,h}{\rm d}\rho = 0\Eq(4.22)$$
which is Maxwell's equal area rule for determining the values of
$\rho_v$ and $\rho_\ell$.

Suppose for all $t\in (0,1)$ $$\psi_{\mu,h}((1-t)\rho_v+ t\rho_\ell)> 
(1-t)\psi_{\mu,h}(\rho_v)+t
\psi_{\mu,h}(\rho_\ell).\Eq(4.23)$$
Suppose also that the infimum in \equ(hat)
is attained exactly at one or two values of $\phi $ when $\rho=\rho_v$ 
and $\rho=\rho_\ell$. Then, depending
whether $\rho_v$ equals or not $\rho_\ell$ and whether there are one or 
two minimizers for $\phi$ in \equ(hat)
there will be one, two, three or four states. To determine if these 
hypotheses hold we must examine a particular
free energy.

\medskip
\medskip
\chap {5. van der Waals Gas.} 5
\numsec= 5
\numfor= 1
\numtheo=1

\bigskip

In this section, we focus on the van der Waals gas for which
$$f_{\beta,\mu_1,\mu_2}(\ro,\rt)=
{1\over\beta}\big[G(\ro)+G(\rt)+D(\rho)\big]+\a\ro\rt-{\s\over
2}(\ro^2+\rt^2)-\mu_1\ro-\mu_2\rt
\Eq(function1)$$
and hence
$$g_{\beta,\mu,h}(\rho,\phi) = {1\over\beta}\big[G\left({\rho\over 
2}(1+\phi)\right)+G\left({\rho\over
2}(1-\phi)\right)+D(\rho)\big]
+{\a\over 4}\rho^2(1-\phi^2)- {\s\over 4}\rho^2(1+\phi^2)
-\mu\rho-h\rho\phi
\Eq(number)$$

  The function $G$, defined in $\R_+$, is assumed to be smooth and moreover

  \item{(1) } $G$ and  $G''$
strictly convex;

\item{(2) } $G'(x)\to -\infty, \quad G''(x)\to +\infty \quad \hbox {for 
} x\to
0.$

The function  $D$ is defined in some interval $(0,b^{-1})$
contained in (and possibly coinciding with) $\R_+$, and is smooth and
strictly convex.
For example, with $G(t) = t\log\ t$ and $D$ given by  \eqv(drho), these
conditions are satisfied. This is the usual two components van der Waals 
gas.

The existence of minimizers follows from the considerations of Section
2, which apply  also in the case of local interactions.
Note however that we have no symmetric monotonicity of the minimizers
and, in general,  no reason to have regularity properties.
Moreover,  the same argument that we used in the proof of Theorem 2.3 shows
that any minimizer satisfies the Euler--Lagrange equations
$$\eqalign{&G'(\ro)+D'(\ro+\rt)+\b\a\rt-\b\s\ro=C_1\cr&
G'(\rt)+D'(\ro+\rt)+\b\a\ro-\b\s\rt=C_2.}\Eq(elzero)$$

We now turn to the study of the structure of
these minimizers. The following theorem gives a criterion for segregation:

\bigskip
\noindent{\bf Theorem 5.1}. {\it With $g$ given by \equ(number), the 
infimum in \equ(hat)  is attained in
exactly one point
$\hat\phi(\rho;\b,\mu,h)$ if $h\neq 0$,
independently of $\r$, or if $h=0$ and
$\b(\s+\a)\le G''(\r/2)$.
If $h=0$ and
$\b(\s+\a)>G''(\r/2)$,
then the infimum is attained in
two points
$\phi_\pm=\pm\hat\phi(\rho;\b,\mu,0)$}
\vskip.3cm

\noindent {\bf Proof:} When there is no ambiguity, we omit subscript 
${\b,\mu,h}$ from
$g_{\b,\mu,h}$.
For fixed values of $\b$, $\mu$ and
$h$ we look for the minimizers of
$g$. To do this, we first fix $\rho$ and look at the solutions of the
equation
$${\partial g\over \partial \phi}(\rho,\phi)=0\Eq(ptphi)$$
for fixed $\rho$. Let $\hat\phi=\hat\phi(\rho;\b,\mu,h)$ be one of the
solutions to
\equ(ptphi). Explicitly, for fixed $\rho>0$, $\hat\phi$ solves the equation
$$H_{\rho}(\hat\phi)=\b(\a+\s)\rho\hat\phi+2h,\Eq(eqphi)$$
with
$$H_\rho(\phi):= G'({\rho\over 2}(1+\phi) )-G'({\rho\over
2}(1-\phi))\Eq(defin)$$
The following properties of the function $H_\rho$ will be relevant
below. Assume $\rho>0$:
\item{}
$$H_\rho(-\phi)=-H_\rho(\phi), \quad H_\rho(0)=0;\Eq(disp)$$
\item{}
$$H_\rho'(\phi)>0, \forall \phi\in (-1,1);\Eq(cres)$$
\item{}
$$H_\rho(\phi)\to \pm \infty, \quad H_\rho'(\phi)\to + \infty\quad
\hbox{ for } \phi\to
\pm 1;\Eq(diver)$$
\item{}
$$\phi H_\r''(\phi)> 0 \quad \hbox{ for } \phi \neq 0, \quad
H''_\rho(0)=0;\Eq(conconv)$$
(Notice that if $G(t)=t\log t$ then $H_\rho(\phi)=2\tanh^{-1}(\phi)$). 
In particular, \equ(conconv) implies that
$$\inf_{\phi\in [-1,1]} H'_\r(\phi)=H'_\r(0)= \r G''({\r\over 2}) \quad
\hbox{ and } H'_\r(\phi)>\r G''({\r\over 2}) \hbox { if }
\phi\neq 0.\Eq(5.11)$$
The above relations are immediate consequences of the assumptions on
$G$: in fact \equ(disp)
just follows from \equ(defin); \equ(cres) follows from
$$H_\r'(\phi)={\rho\over 2}G''({\rho\over 2}(1+\phi) )+G''({\rho\over
2}(1-\phi))\Eq(5.12)$$
and the fact that $G$ is strictly convex. \equ(diver) follow from
condition (2) on $G$  and finally,
\equ(conconv) follows from
$$H_\r''(\phi)={\rho^2\over 4}G'''({\rho\over 2}(1+\phi)
)-G'''({\rho\over 2}(1-\phi))\Eq(5.13)$$
and the strict monotonicity of $G'''$.

We have
$${\partial^2 g\over \partial\phi^2}(\r,\phi)= {\rho\over
2}\left(H_\r'(\phi)-(\a+\s)\b\r\right)\ge {\rho^2\over
2}\left(G''({\r\over 2})-(\a+\s)\b\right).\Eq(derv2)$$
Therefore, if
$$\b(\a+\s)< G''({\r\over 2}),\Eq(notran)$$
the r.h.s. of \equ(derv2) is positive, the function $g(\r,\phi)$, for
$\r$ fixed is
a strictly convex function of
$\phi$ and hence it has a unique minimizer solving
$\hat\phi(\rho;\b,\mu,h)$ solving \equ(eqphi),
because of condition (2) on $G$ permits to exclude that it is a corner
solution.
On the other hand, if
$$\b(\a+\s)>G''({\r\over 2}),\Eq(tran)$$
by \equ(conconv), there is an interval $(-\phi_s,\phi_s)$ where the
function $g(\r,\phi)$,
for $\r$ fixed is concave while in the complement it is convex.  The
value $\phi_s$ is
determined by the condition
$$H_\r'(\phi_s)=(\a+\s)\b\r.\Eq(5.17)$$
Therefore, it is possible to get more than one stationary point, satisfying
\equ(eqphi).

We assume that \equ(tran) is verified and look at the stationary points.
We distinguish
two cases: $h=0$ and $h\neq 0$.

We assume first $h=0$. Then, clearly,
$\phi=0$ solves
\equ(eqphi). The condition \equ(tran) ensures that $g(\r,\phi)$ is
concave in
$\phi=0$ and hence $\phi=0$ is not a minimizer. Moreover, by the
symmetry, if
$\hat\phi(\rho;\b,\mu,0)$ solves \equ(eqphi) the same is true for
$-\hat\phi(\rho;\b,\mu,0)$. \equ(cres) and
\equ(conconv) show that there is exactly one positive value
$\hat\phi(\rho;\b,\mu,0)\in
(\phi_s,1)$ solving \equ(eqphi). Consequently, we have the two
stationary points
$$\phi_\pm=\pm \hat\phi(\rho;\b,\mu,0).\Eq(5.18)$$

Now we take $h\neq 0$. One immediately realizes that there is $\bar
h(\rho)$ such
that, if
$h\notin [-\bar h(\r),\bar h(\r)]$ then there is only one solution to
\equ(eqphi), for
$h=\pm \bar h(\r)$ there are two solutions, one of them in
$(-\phi_s,\phi_s)$ and the
other one is the minimizer. Finally, if $h\in (-\bar h(\r),\bar h(\r))$
there are three
solutions, one in $(-\phi_s,\phi_s)$ and the others, in the complement,
corresponding to
local minimizers; moreover they have different signs. Let
$\hat\phi_1(\rho;\b,\mu,h)>0$ and $\hat\phi_2(\rho;\b,\mu,h)<0$
denote the two local minimizers. We show that
$g(\rho,\hat\phi_1(\rho;\b,\mu,h))\neq
g(\rho,\hat\phi_2(\rho;\b,\mu,h))$ and hence there is only one absolute
minimizer.
To show this, we compare with the case $h=0$, where
$\hat\phi_2(\rho;\b,\mu,0))=-\hat\phi_1(\rho;\b,\mu,0))$ and the
corresponding values of
$g$ are equal. We define
$$J_i(h)=g_{\b,\mu,\h}(\rho,\hat\phi_i(\rho;\b,\mu,h)), \quad 
i=1,2.\Eq(5.19)$$
Since $\hat\phi_i$ are stationary points for $g$, we have
$$\eqalign{&{d\over d h}J_i= {\partial g_{\b,\mu,\h}\over \partial \phi}
(\rho,\hat\phi_i(\rho;\b,\mu,h)){\partial
\hat\phi_i \over \partial h}+ {\partial
g_{\b,\mu,\h}\over \partial h}
(\rho,\hat\phi_i(\rho;\b,\mu,h))\cr&={\partial
g_{\b,\mu,\h}\over \partial h} (\rho,\hat\phi_i(\rho;\b,\mu,h))=-\r
\hat\phi_i(\rho;\b,\mu,h).}\Eq(5.20)$$
Therefore, for $h>0$
$$J_1(h)<J_1(0)=J_2(0)<J_2(h)\Eq(5.21)$$
and {\it vice--versa} for $h<0$. \eop

\bigskip

Next we give a criterion for condensation-evaporation transition. Such a 
transition occurs exactly for those
values of $\a,\sigma,\b,\mu,h$ for which
  the function
$\psi_{\b,\mu,h}(\r)$ fails to be strictly convex. Our first result 
pertaining to this is the following:

\medskip

\noindent{\bf Lemma 5.2} {\it  Let $\hat \phi(\rho)$ be any  minimizer
of $g(\rho,\phi)$ in \equ(number) with respect to $\phi$, so that 
$\psi(\r)=g(\r,\phi(\r))$. Then  $\psi''(\rho)$
is strictly positive if  the point
$(\rho,
\hat\phi(\rho))$ is in the regions where  the Hessian of $g$, $D^2g$, is 
positive definite. }

\medskip

\noindent {\bf Proof.}

We have:
$$\psi'= {\partial g\over \partial\r} + {\partial g\over
\partial\f}\hat\f'\ ,\Eq(5.22)$$
$$\eqalign{\psi''&= {\partial^2 g\over \partial\r^2} + 2{\partial^2 g\over
\partial\f\partial\r}\hat\f'+{\partial^2 g\over
\partial\f^2}(\hat\f')^2+ {\partial g\over
\partial\f}\hat\f''\cr&=
\langle (1,\hat\f'), D^2 g(1,\hat\f')\rangle +{\partial g\over
\partial\f}\hat\f'',}\Eq(hesscomp)$$
$D^2g$ being the Hessian matrix of $g$  and $\langle\,\cdot\,
,\,\cdot\,\rangle$ the usual
scalar product in $\R^2$.

When $\hat\f(\r)$ is a minimizer, above relation reduces to
$$\psi''= \langle (1,\hat\f'), D^2 g(1,\hat\f')\rangle\Eq(hess)$$
Therefore, $\psi''(\rho)$ is strictly positive if the point $(\rho,
\hat\phi(\rho))$ is in the regions where $D^2g$ is positive definite.\eop
\bigskip
The above result can be restated in terms of the function $f$. Since 
$\hat\phi(\r)$ solves \equ(eqphi), it is
plain  that
$\psi''(\rho)=\langle j^T(1,\hat\f'), D^2 f j^T(1,\hat\f')\rangle$
where
$$j=\left( \matrix{
\displaystyle{{\partial \r\over \partial\ro}} &\displaystyle{{\partial
\r\over \partial \rt}}
\cr
\displaystyle{{\partial \phi\over \partial\ro}} &\displaystyle{{\partial
\phi\over \partial \rt}}
}\right).\Eq(5.25)$$
Therefore we study the positivity properties of $D^2 f$.
Its expression is:
$$D^2f=\left( \matrix{
\displaystyle{G''(\ro)-\b\s+D''(\r)} &\displaystyle{D''(\r)+\b\a}
\cr
\displaystyle{D''(\r)+\b\a} &\displaystyle{G''(\rt)-\b\s+D''(\r)}
}\right).\Eq(5.26)$$
As a consequence of Lemma 5.1, when the Hessian of $f$ is positive 
definite, there is no
condensation-evaporation transition.
However, if $\s$ is sufficiently large, the diagonal terms may become
negative, while for
$\a$ sufficiently large the off diagonal terms may make $\det(D^2 f)<0$.

We now consider explicitly some examples for the case $G(t)= t\ \log \ t$.

\item{1)}
  $\s=0$ and $D=0$ (ideal gas).
Then $D^2f$ is positive definite  in the region ${\cal P}$ defined by
$${\cal P}=\{(\ro,\rt)\,|\, \ro\rt<\a^{-2}\b^{-2}\}.\Eq(cond12)$$
To check this, first suppose $\alpha\beta\rho > 2$. With our choice of 
$G$, \equ(eqphi) reduces to
$$\phi = \tanh\left({\alpha\beta\phi\rho\over 2}\right).\Eq(consistency)$$
With $\xi = \alpha\beta\rho/2$ it becomes
$$\phi = {\tanh}(\xi \phi)\ .\Eq(concon)$$
For all strictly positive values of $\xi\phi$,
$${\sinh}(\xi\phi) > \xi\phi\ .\Eq(5.30)$$
By \eqv(concon) we can divide the left hand side by ${\tanh}(\xi
\phi)$ and the right hand side
by $\phi$ without affecting the inequality.
It follows that
${\cosh}(\xi\phi) > \xi$, and hence that $(1 - {\tanh}^2(\xi\phi))^{-1} 
 > \xi^2$.
By \eqv(concon), this means that
$$(1 - \phi^2)^{-1} \ge \left({\alpha\beta\rho\over
2}\right)^2\Eq(sepcon)$$
whenever $\alpha\beta\rho > 2$. Of course if $\alpha\beta\rho \le 2$, 
then, by Theorem 5.1
$\phi = 0$,
and so \eqv(sepcon) holds in all cases, and there is equality if and only if
$\alpha\beta\rho =2$.
Since
$$\ro\rt={\r^2\over 4}(1-\phi^2)\le {1\over \alpha^2\beta^2},\Eq(5.31)$$
by \equ(concon), the condition in \equ(cond12) is fulfilled for all the 
values of the parameters but for
$\alpha\beta\rho =2$. In this case however, $\phi=0$ and by 
\equ(hesscomp) $\psi''={2\over\rho}>0$. Hence
$\psi(\r)$ is strictly convex and there is
no condensation-evaporation transition for any value of the parameters.

\item{2)} $D(\r)$ is given  by \equ(drho) and  $\s=0$.
  Then the strict convexity of $G$ and $D$ ensures that the diagonal terms
are positive and
we only need to check  $\det(D^2 f)>0$. Hence, in this case $D^2f$ is
positive definite in
$${\cal Q}=\{(\ro,\rt)\,|\, \det(D^2 f)>0\}\Eq(5.33)$$
and its boundary,
$$\partial{\cal Q}=\{(\ro,\rt)\,|\, \det(D^2 f)=0\}$$
separates it from the region where $D^2f$ is not positive definite.
Explicitly, the condition $\det(D^2 f)>0$ can be written
$${1\over \ro\rt}-\a^2\b^2 + D''(\r)({1\over \ro}+{1\over 
\rt}-2\a\b)>0.\Eq(5.34)$$
Note that the set ${\cal Q}$ contains the set
$${\cal P}=\{(\ro,\rt)\,|\,{1\over \ro\rt}-\a^2\b^2>0\}\Eq(5.35)$$
because the arithmetic mean of two positive numbers is not less than the
geometric mean.
Since the equation for $\phi$ is unaffected by the presence of $D$, 
$\psi(\rho)$ is strictly convex
and we get the same conclusion as before.

\item{3)} If $\s>0$, the set ${\cal P}$ is replaced by
$${\cal P}_\s=\{(\ro,\rt)\,|\,({1\over \ro}-\b\s)({1\over
\rt}-\b\s)-\a^2\b^2>0\}.\Eq(5.36)$$

Therefore, as $\sigma$ increases the set ${\cal P}_\s$ shrinks thus 
encouraging the condensation-evaporation transition.
\medskip
Korteweg [\rcite{K}] has also discussed a situation with segregation and 
four
phases in equilibrium in a mixture of
two van der Waals gases.
Korteweg's paper concerns the $\a<0$
case where there is attraction between different species too, and to
which rearrangement arguments do not apply, though  segregation is still 
possible because it is controlled by
$\s+\a$.

Up to now we just gave general conditions for the absence of
condensation-evaporation transition.
We now derive a formula for the pressure which allows us to draw 
conclusions
about the not strict convexity of the function $\psi$.
Introduce the specific volume
${v}=\r^{-1}$ and define the function
$$q(v)=v\psi(v^{-1}),\Eq(5.37)$$
which has the same convexity properties of $\psi$ in the corresponding
points.
A simple calculation shows that,
$$q(v)=-\log(v-b) -{\b \tilde\sigma(\hat\phi(v^{-1}))\over 4 v}
+\gamma(\hat\phi(v^{-1}))\Eq(5.38)$$
with
$$\gamma(\hat\phi)= {1\over 2} \log\left({1-\hat\phi^2\over 4}\right)
+{\hat\phi\over 2}\log\left({1+\hat\phi\over 
1-\hat\phi}\right)-h\hat\phi-\mu\Eq(5.39)$$
and
$$\tilde\sigma(\hat\phi)=\sigma-\a+\hat\phi^2(\sigma+\a).\Eq(sigmatilde)$$
Now, by \equ(eqphi), we have
$${\b \tilde\sigma'\over 4 v} +\gamma'=0.\Eq(5.41)$$
Hence  the pressure $p$ as function of the specific volume is given by
$$\b p(v):=-q'(v)={1\over v-b}-{\b \tilde\sigma(\hat\phi(v^{-1}))\over 4 
v^2}.\Eq(press)$$
This expression is similar to that for the one component van der Waals
gas, but for the dependence of $\tilde\sigma$ on $v$
through $\hat\phi$.

Several cases are possible, depending on $\a$ and $\s$. Consider for
example the case $\a >\s$. Then, in the absence of segregation,
the pressure is a strictly decreasing function of $v$. Since the
pressure must be equal in the liquid and vapor phases,
this precludes a condensation-evaporation transition in the absence of
segregation. (This is quite different from the usual van der Waals
case, or the case considered by Korteweg). The segregation permits a
condensation-evaporation transition, and lowers the pressure in
the liquid phase from what it would be without segregation. As the
temperature is lowered further,
the vapor phase may also undergo segregation. In this situation, the
volume fractions are not determined. However, in the final section
of the paper we shall see that because the local interaction model is
the large volume limit of a model in which the minimizers
must be symmetric decreasing, we can say more about this case. Thus when
$\a >\s$, we have three critical inverse temperatures
$\b_1 < \b_2 < \b_3$. The first corresponds to a segregation transition.
At the second there is a condensation-evaporation transition
at which a segregated liquid phase is produced, and an unsegregated
vapor phase. At the third, the vapor phase as well segregates.
As a consequence of the above formula for the pressure, we prove the 
following:
\bigskip
\noindent{\bf Theorem 5.3}. {\it
Suppose that $\a\le\sigma$. Then there is a critical inverse temperature 
$\beta_c$ such that for
$\beta>\beta_c$ a condensation-evaporation transition occurs, with or 
without segregation, depending on the
values of $\beta$, $h$ and $mu$, according to Theorem 5.1. On the other 
hand, if $\a>\sigma$, then either there
is a unique minimizer or the condensation-evaporation  happens at lower 
temperature than segregation.}
\bigskip
\noindent{\bf Proof:} First we consider the case $\a<\sigma$. We 
differentiate \equ(press) and  get
$$\b p'(v)=-{1\over (v-b)^2}+{\b \tilde\sigma(\hat\phi(v^{-1}))\over 2 
v^3}+{\b
\hat\phi(\sigma+\a)\hat\phi'(v^{-1})\over 2 v^4}.\Eq(press1)$$
Since $\hat\phi$ is non decreasing, the last two terms are both positive 
and at least linearly increasing with
$\beta$. Hence, for any fixed $v$ and sufficiently large $\beta$ they 
dominate the first term. On the other hand,
for any fixed $\beta$ and $v$ sufficiently close to $b$, the first term 
dominates. Therefore the pressure is not
monotone.

Let us now consider the case $\a >\s$. In the absence of segregation,
the pressure is a strictly decreasing function of $v$ and
this precludes a condensation-evaporation transition in the absence of
segregation. (This is quite different from the usual van der Waals
case, or the case considered by Korteweg). The segregation permits a
condensation-evaporation  transition, and lowers the pressure in
the liquid phase from what it would be without segregation. As the
temperature is lowered further,
the vapor phase may also undergo segregation. \eop
\bigskip
In conclusion we may have at most four phases, characterized by the 
following values of the density:
$$\r^\pm_v=\r_v{1\pm\hat\phi(\r_v)\over 2}, \quad
\r^\pm_l=\r_l{1\pm\hat\phi(\r_l)\over 2}.\Eq(5.44)$$

\bigskip
We now return to the search  for the constrained minimizers which are
density profiles $\r_1(x),\r_2(x)$ such that
$$\int_{\L} dx \r_1(x)=n_1|\L|, \quad \int_{\L} dx 
\r_2(x)=n_2|\L|.\Eq(5.45)$$

Clearly if $\min\{n_1,n_2\}\le\r^-_v$ or $\max \{n_1,n_2\}\ge\r^+_l$ the
only possible solutions are $\r_1=n_1$ and $\r_2=n_2$.
Otherwise, non homogeneous minimizers are possible. Because of the
previous lemma it is clear that $\r_1$ and $\r_2$ can only take the
above values in four regions $A$, $B$, $C$ and $D$ whose volumes are the
only relevant properties because we are dealing with the
local interaction case.   The minimizer $(\r_1^\star,\r_2^\star)$  is
given by
$$\r^\star_1(x)=\cases{\r^+_l \quad \hbox{ if } x\in A,
\cr
                   \r^-_l \quad \hbox{ if } x\in B,
\cr
                   \r^+_v \quad \hbox{ if } x\in C,
\cr
                   \r^-_v \quad \hbox{ if } x\in D,}
\quad\quad
    \r^\star_2(x)=\cases{\r^-_l \quad \hbox{ if } x\in A,
\cr
                   \r^+_l \quad \hbox{ if } x\in B,
\cr
                   \r^-_v \quad \hbox{ if } x\in C,
\cr
                   \r^+_v \quad \hbox{ if } x\in D.}\Eq(minzeroran)$$
                   or
viceversa.
Let $a$, $b$, $c$ and $d$ denote the ratios of
above volumes with $|\L|$. The constraints are then
$$a\r^+_l+b\r^-_l+c \r^+_v+d\r^-_v= n_1,\quad a\r^-_l+b\r^+_l+c
\r^-_v+d\r^+_v= n_2,\Eq(5.47)$$
which, together with the relation
$$a+b+c+d=1\Eq(5.48)$$
do not suffice to determine the fractions occupied by each phases.
Note that, in the case $\b_s(\r_l)<\b< \b_s(\r_v)$ we have $\r_v^+=\r_v^-$
and the relations available are sufficient to determine the volume
fractions occupied by the three phases.
\bigskip
Now we return to the finite volume case where the rearrangement 
inequalities take over and ensure regularity of
the minimizers.  In fact because the minimizers must be symmetric monotone,
not all of the phases can be in contact,
and there are constraints on the ordering of the densities.
Suppose for example that we have
$$\rho_\ell^+ >  \rho_\ell^- > \rho_v^+  > \rho_v^-\ .\Eq(5.49)$$
We shall show that this is impossible.

Pick numbers $a$, $b$ and $c$ separating these density values so that
$$\rho_\ell^+ > a >  \rho_\ell^- > b >  \rho_v^+  > c > \rho_v^-\ 
.\Eq(5.50)$$
Define
$$\eqalign{ A &= \{\ x\ : \rho_1(x) > a\ \}\qquad B =\{\ x\ : a \ge
\rho_1(x) > b\ \}\cr
\qquad C &=\{\ x\ : b \ge  \rho_1(x) > c\ \}\qquad
D = \{\ x\ : c \ge \rho_1(x) \ \}\ .\cr}$$
By the rearrangement inequalities, we see that $\rho_2$ must take its
minimum in $A$, but we know from \eqv(minzeroran)
that $\rho_2$ must take the value $\rho_\ell^-$ in $A$. So we must have
instead that
$$\rho_\ell^+ >  \rho_v^+ > \rho_v^-  > \rho_\ell^-\ .\Eq(5.51)$$
It is not evident how to deduce this in such generality by direct
consideration of the local interaction model.
The  above regions are disjoint. To further determine the structure we 
need to consider the surface tension
across their boundaries. In a later paper we will rigorously investigate 
the surface tension. Here we proceed on
the assumption that because of it the phase boundaries will be so 
arranged as to have minimum surface area.
  Then, for a two dimensional
torus,  these regions would be arranged in the manner indicated in the 
diagram below:
\bigskip\bigskip\bigskip
\phantom{.}

\vskip 2.5 true in
\includegraphics{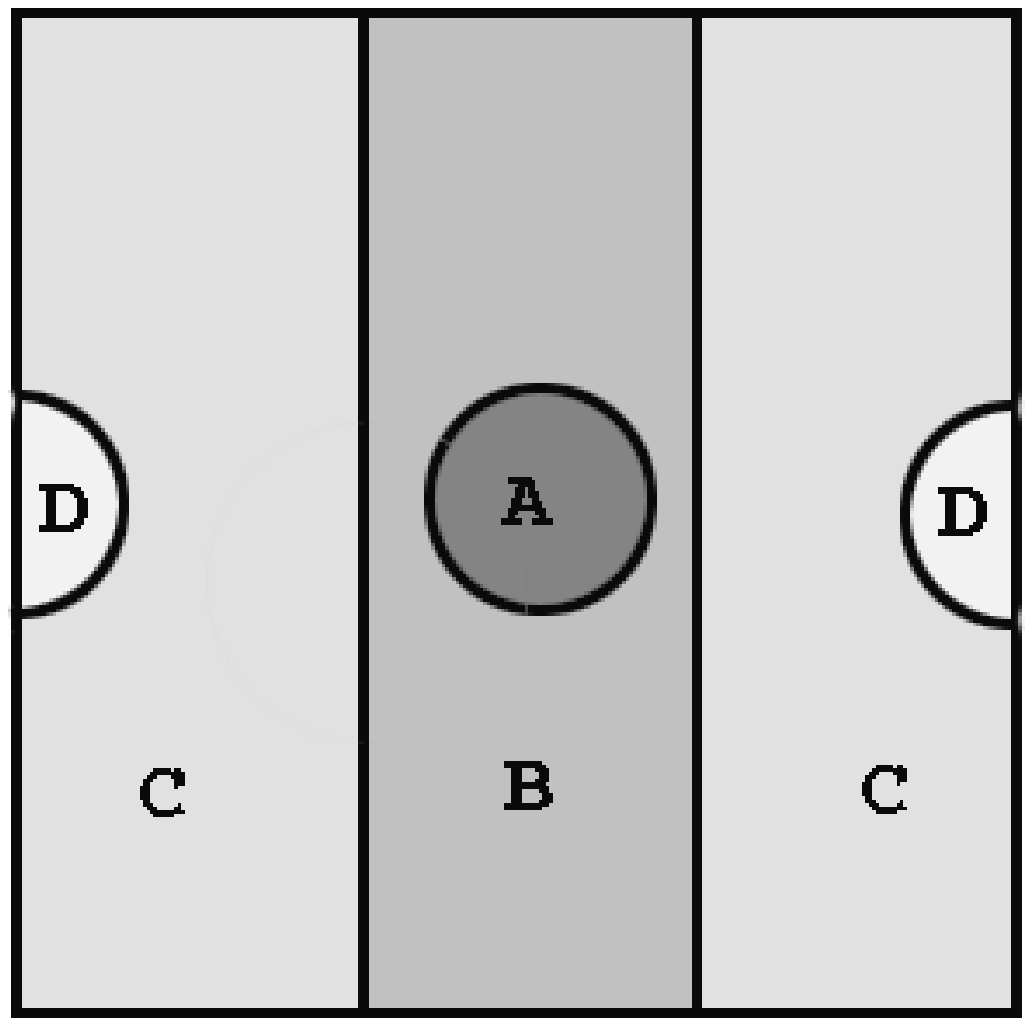}

Note that on a torus, the solutions of the isoperimetric problem are
either disks, or
bands depending on the size of the area to be enclosed.

Now we see that the liquid state rich in species one  can only be in
contact with the  vapor state rich in species one,
which in turn can only be in contact with the species one rich liquid
state and the species 
two poor vapor state.
In the case that all four phases are present, it is the surface tension
of the domain boundaries that
fixes the volume fractions in the finite range case.
\medskip\medskip
\medskip
\chap {6. The Microcanonical Minimization Problem.} 6
\medskip
\numsec= 6
\numfor= 1
\numtheo=1
\medskip
In the previous sections we have considered a minimization problem 
related to a model
in thermal contact with a reservoir. In this section we extend our 
results to a kinetic  model
preserving  energy, but, for sake of simplicity, we confine ourselves to 
the ideal gas
case with  repulsion between different species and no attractive part. 
For example, consider  the
Boltzmann-Vlasov equations  in [\rcite{BELM}], for the position and velocity
distributions $f_i(x,v,t)$. The results on the minimization problem we have
considered  in the previous sections can
be applied to determine the equilibrium solutions of these equations.

The equilibrium solutions of such a model are the minimizers of the 
entropy with energy
and mass constraint.
Therefore, the variational problem we want to deal with is the 
following: find the
minimizers of
$${\cal S}(\bf f)=\int {\rm d}v\int_\Lambda {\rm d}x \sum_{i=1}^2 
f_i\log f_i\Eq(6.1)$$
in the set
$${\cal D}_{e,n_1,n_2}=\{\bf f=(f_1,f_2) \,|\, |\Lambda|^{-1}\int {\rm 
d}v \int_\Lambda
{\rm d}x f_i= n_i,\quad i=1,2;\quad E(\bf f)=e\}\Eq(6.2)$$
where $f_1$ and $f_2$ are probability densities in the phase space 
$\Lambda\times \R^3$ and
$$E\bf (f)= |\Lambda|^{-1}\left[\int {\rm d}v \int_\Lambda {\rm d} x 
{v^2\over 2} (f_1+f_2)
+
\int_\Lambda {\rm d}x \rho_{f_1} V*\rho_{f_2}\right], 
\quad\rho_{f_i}=\int {\rm d}v
f_i.\Eq(6.3)$$
This is equivalent to minimizing without constraints the functional
$${\cal G}(\bf f)= {\cal S}(\bf f)+ \b E\bf (\bf f) -\sum_{i=1}^2 \mu_i 
\int dv
\int_\Lambda dvf_i\Eq(6.4)$$
on
$${\cal D}=\{\bf f=(f_1,f_2) \,|\, |\, f_i\in L_1^+(\Lambda\times 
\R^3)\}\Eq(6.5)$$
with $\b$, $\mu_i$, Lagrange multipliers to be determined by the 
constraints.

As  is well known, the Euler-Lagrange equations for this functional 
force the
$v$-dependence of $f_i$ to be a Maxwellian at inverse temperature $\b>0$ 
and hence we
are reduced to the following variational principle:
$${\cal F}(\rho_1,\rt)=\int_\Lambda {\rm d}x \sum_{i=1}^2 \rho_i[\log 
\rho_i -\bar\mu_i]+
\b\left[ \int_\Lambda {\rm d}x\rho_1 V*\rho_2,\right]\Eq(free)$$
on
$$\overline{{\cal D}}=\{(\rho_1,\rho_2) \,|\, | \rho_i\in 
L_1^+(\Lambda)\},\Eq(6.7)$$
with $\b>0$, $\bar \mu_i$ Lagrange multipliers to be determined by the 
constraints
$$\overline{E}\bf (\rho_1,\rho_2)= |\Lambda|^{-1}\left[ \int_\Lambda 
{\rm d} x {3\over 2}
\b^{-1}(\rho_1+\rho_2) + \int_\Lambda {\rm d}x \rho_1 V*\rho_2\right]=e, 
\quad
|\Lambda|^{-1}
\int_\Lambda {\rm d}x \rho_i= n_i.\Eq(6.8)$$
Here $\bar \mu_i=\mu_i+\log (\sqrt{2\pi \b^{-1}})^3$.
This is equivalent to minimizing
$$\overline{\cal S}(\rho_1,\rho_2)=\sum_{i=1}^2\int_\lambda {\rm d}x 
\rho_i\log
\rho_i\Eq(6.9)$$
on the set
$$\overline{{\cal D}}_{e,n_1,n_2}=\{\bf (\rho_1,\rho_2) \,|\, |\Lambda|^{-1}
\int_\Lambda {\rm d}x \rho_i= n_i,\quad i=1,2;\quad \bar E(\bf 
f)=e\}.\Eq(6.10)$$

The discussion in previous sections obviously extends to the study of 
the minimizers of the functional
\equ(free). Therefore we now only deal with the solvability of the 
conditions on the
Lagrange multipliers. This relies essentially on the local interaction 
case, because the arguments of section 3
apply.

Consider first  the local interaction  case, with unconstrained energy. 
For $\b$ sufficiently small,
($\a\b n<2$) there is only a homogeneous solution. The energy as a 
function of $\b$
is given by
$$E(\b)={3\over 2} \b^{-1}n+ \a n^2|\Lambda|,\Eq(6.11)$$
i.e. the usual linear behavior in $\b^{-1}=T$.
For sufficiently large $\b$ ($\a\b n>2$) the solution is of
the form:
$$\rho_1(x)=\cases{\rho_+, \quad x\in A,\cr \rho_- \quad x\in 
\Lambda-A,}\quad
\rho_2(x)=\cases{\rho_-, \quad x\in A,\cr \rho_+ \quad x\in 
\Lambda-A,}\Eq(6.12)$$
with $\rho_\pm=\rho(1\pm \phi(\b))/2$, $\phi(\beta)$ the unique positive 
solution to
$$\tanh^{-1} \phi(\b)= {\a\b\rho\over 2}\phi(\b),\Eq(6.13)$$
and $A$ a suitable subset of $|\Lambda|$, whose volume is determined to 
fulfill the
mass constraints.  Of course in the present situation $\rho=n=n_1+n_2$ 
and hence the
energy is given by
$$E(\b)={3\over 2} \b^{-1}n+ \a \int_\Lambda{\rm d} 
x\rho_1\rho_2={3\over 2} \b^{-1}n+ \a
|\Lambda|\rho_+\rho_-={3\over 2} \b^{-1}n+ {\a|\Lambda|n^2\over 
4}(1-\phi(\b)^2).\Eq(6.14)$$
This is obviously continuous  because $\phi(\b)\to 0$ as $\b\to 2/n\a$.
Moreover by using arguments similar to those of section 5 one can show that
$\phi(\b)$ is increasing for $\b>2/n\a$. Therefore $E(\b)$ is monotone 
decreasing,
as sum of two decreasing functions, and hence invertible.

The function $\beta(e)$ is well defined and, with this definition, the 
phase space densities for the two
species are given by
$$f_i(x,v)= \r_i(x) {{\rm e}^{-\beta(e)v^2/2}\over (2\pi 
\beta(e)^{-1})^{3/2}},\Eq(effe)$$
where $\r_i$ are the minimizers of the free energy at temperature 
$\beta(e)$.

These densities give us the minimizers of the entropy ${\cal S}(f)$ 
under the constraints $E(f)=e$.  Note that
the determination of the function $\beta(e)$ in \equ(effe) require 
knowledge of the partition of $e$ into its
kinetic and interaction parts. This is provided by the results in the 
previous sections.
\bigskip

\noindent {\bf Acknowledgments.}  We thank E. Presutti and A.
Levelt Sengers for helpful comments.  We also thank Anneke
Sengers for supplying us with the papers of Korteweg.  The
authors would like to thank the kind hospitality of IHES, France, where
their collaboration on this problem started.  The work of E.C. and
M.C.C. was supported by U.S. N.S.F. grant DMS 00-70589, J.L.L. was supported by AFOSR Grant AF
49620-01-1-0154, and NSF Grant DMR 98-13268. R.M. partially supported by 
MIUR, Cofin-2000 and INFM.

\bigskip
\medskip
\noindent{\bf Appendix A.}
\numfor= 1

\medskip
Here we prove the rearrangement inequalities used in the paper.
\medskip
\noindent{\bf Proof of Lemma 2.2}
We shall apply a sequence of symmetrization operations, and show that unless
the minimizer is
symmetric monotone, then one of the these operations would strictly
lower the integrals in \eqv(reran1) and \eqv(reran2).

These symmetrization operations  are ``transplants'' to the
torus of the
following symmetrization operation on the circle
${\cal S}^1 = \{(x_1,_2)\ |\ x_1^2 + x_2^2 =1\}$ in the $x_1,x_2$ plane.
Fix any
unit vector
${\bf u}$ on ${\cal S}^1$, and define the reflection
$R_{ u}$ on ${\cal S}^1$ given by
$$R_{\bf u}x = {\bf x} - 2({\bf u}\cdot {\bf x}){\bf u}\Eqa(A.1)$$
where ${\bf x} = (x,y)$, and the dot denotes the usual inner product.
It is also convenient to define $R_{{\bf u}}\theta$ for $-\pi < \theta <
\pi$ by
$$R_{{\bf u}}(\cos(\theta),\sin(\theta)) = (\cos(R_{{\bf
u}}\theta),\sin(R_{{\bf u}}\theta))\ .\Eqa(A.2)$$
Next, define an operator, denoted $R_{\bf u}^+$, on measurable functions
on ${\cal S}^1$
by
$$R^+_{\bf u}g(x) =
\cases{ \max\{g({\bf x}),g(R_{\bf u}{\bf x})\} & if ${\bf x}\cdot{\bf
u_0}\ge0$\ ,
\cr \min\{g({\bf x}),g(R_{\bf u}{\bf x})\} & if ${\bf x}\cdot{\bf
u_0}<0$\ \cr}
\Eqa(zoodah)$$
where ${\bf u}_0$ denotes $(0,1)$. That is, the line through the origin
perpendicular to
${\bf u}$ divides the circle in two, and the two halves are identified
by the reflection
fixing this line. The symmetrization operation swaps values at reflected
pairs of points,
if necessary, so that the large value is always on the side containing
the ``north
pole'', $(0,1)$. We also define an operator $R_{\bf u}^-$ in the same
manner, except that
we put the small values on the side with the ``north pole''. In other
words, we
interchange the minimum and maximum in \eqv(zoodah)

We now state the lemma for which these definitions were made.
\bigskip
\noindent{\bf Lemma A.1} {\it  Let $F$ be a symmetric function on
$\R_+\times \R_+$ that
satisfies \eqv(cheddar1) everywhere on its domain.
Let $K$ be any strictly decreasing
non-negative function
on $\IR_+$. Then for any two bounded non--negative measurable functions
$g$ and $h$ on
${\cal S}^1$ identified with $[-\pi,\pi)$,
$$\int_{-\pi}^\pi \int_{-\pi}^\pi g(\theta)K(|\theta -\phi|)h(\phi){\rm
d}\theta{\rm d}\phi
\ge
\int_{-\pi}^\pi \int_{-\pi}^\pi R_{\bf u}^+g(\theta)K(|\theta -\phi|)R_{\bf
u}^-h(\phi)){\rm d}\theta{\rm d}\phi \Eqa(rear)$$
and
$$\int_{-\pi}^\pi  F( g(\theta), h(\theta ){\rm d}\theta
\ge
\int_{-\pi}^\pi F(R_{\bf u}^+g(\theta), R_{\bf u}^-h(\theta)){\rm
d}\theta\ . \Eqa(A.5)$$
Moreover,  there is equality in inequality \eqv(rear) if and
only if for some
fixed $\theta_0$,
$R_{\bf u}^+g(\theta - \theta_0)=g(\theta)$ and
$R_{\bf u}^-h(\theta - \theta_0)=h(\theta)$ for almost all $\theta$.}
\bigskip

Lemma A.1 will be  proved after using it to prove Lemma 2.2. The
argument is
adapted from  [\rcite{BT}], which however does not consider cases of
equality.

Consider first \eqv(reran1).
Fix any index $i$, and fix values of $x_j$ and
$y_j$ for $j\ne i$. Determine a sequence of unit vectors $\{{\bf u}_j\}$
inductively as follows. Suppose that
the ${\bf u}_i$ have been determined for $i<j$. For $i<j$, define $g_i$
and $h_i$ inductively by
$$g_{i} = R_{{\bf u}_i}^+g_{i-1}\ ,\qquad g_0 = g\qquad{\rm and}\qquad
h_{i} = R_{{\bf u}_i}^-h_{i-1}\ ,\qquad h_0 = h\ .\Eqa(A.6)$$
Now choose ${\bf u_j}$ so that
$$\eqalign{
&\int_{-\pi}^\pi \int_{-\pi}^\pi R_{{\bf u}_j}^+g_{j-1}(\theta)K(|\theta
-\phi|)
R_{{\bf u}_j}^-h_{j-1}(\phi)){\rm d}\theta{\rm d}\phi -
\int_{-\pi}^\pi \int_{-\pi}^\pi g_{j-1}(\theta)K(|\theta
-\phi|)h_{j-1}(\phi)){\rm d}\theta{\rm d}\phi\cr
&\le
{1\over 2}\sup_{\bf u}\left\{\left(
\int_{-\pi}^\pi \int_{-\pi}^\pi R_{\bf u}^+g(\theta)K(|\theta -\phi|)R_{\bf
u}^-h(\phi)){\rm d}\theta{\rm d}\phi -
\int_{-\pi}^\pi \int_{-\pi}^\pi g_{j-1}(\theta)K(|\theta
-\phi|)h_{j-1}(\phi)){\rm d}\theta{\rm d}\phi\right)\right\}\ .\cr}$$
That is, we choose the direction ${\bf u}_j$ to get an effect that is at
least half of the largest possible effect
at that stage.

The operations $R_{{\bf u}_j}^\pm$ preserve the modulus of continuity,
so if $g$ and $h$ are continuous, the
sequences $\{g_j\}$ and $\{h_j\}$ are strongly compact. Passing to a
subsequence along which limits exist,
define
$$g^\star = \lim_{j\to\infty}g_j\qquad{\rm and}\qquad  h_\star =
\lim_{j\to\infty}h_j\ .\Eqa(A.7)$$
It follows from the choice of the sequence that for every ${\bf u}$,
$$\int_{-\pi}^\pi \int_{-\pi}^\pi R_{{\bf
u}_j}^+g^\star(\theta)K(|\theta -\phi|)
R_{{\bf u}_j}^-h_\star(\phi)){\rm d}\theta{\rm d}\phi =
\int_{-\pi}^\pi \int_{-\pi}^\pi g^\star(\theta)K(|\theta
-\phi|)h_\star(\phi)){\rm d}\theta{\rm d}\phi\ .\Eqa(A.8)$$
By Lemma 5.3, this is only possible if $g^\star$ is symmetric monotone
about the ``north pole'',
and $h_\star$ is symmetric monotone about the south pole, $-{\bf u}_0$.
Thus, since the $\{g_j\}$ are equimeasurable, as are the
$\{h_j\}$, $g^\star$ and $h_\star$ are the symmetric monotone
rearrangements of $g$ and $h$ respectively. Since
$$\int_{-\pi}^\pi \int_{-\pi}^\pi R_{{\bf
u}_j}^+g_{j-1}(\theta)K(|\theta -\phi|)
R_{{\bf u}_j}^-h_{j-1}(\phi)){\rm d}\theta{\rm d}\phi -
\int_{-\pi}^\pi \int_{-\pi}^\pi g_{j-1}(\theta)K(|\theta
-\phi|)h_{j-1}(\phi)){\rm d}\theta{\rm d}\phi \le 0\Eqa(A.9)$$
for each $j$, it follows that
$$\int_{-\pi}^\pi \int_{-\pi}^\pi g^\star(\theta)K(|\theta -\phi|)
h_\star(\phi)){\rm d}\theta{\rm d}\phi \le
\int_{-\pi}^\pi \int_{-\pi}^\pi g(\theta)K(|\theta -\phi|)h(\phi)){\rm
d}\theta{\rm d}\phi\ .\Eqa(A.10)$$
The requirement of continuity is easily removed by a density argument,
exactly as in   [\rcite{BT}]. This proves the
inequality in general. Now for equality to hold, it must be the case that
$$ \int_{-\pi}^\pi \int_{-\pi}^\pi R_{{\bf u}_j}^+g(\theta)K(|\theta -\phi|)
R_{{\bf u}_j}^-h(\phi)){\rm d}\theta{\rm d}\phi  = \int_{-\pi}^\pi
\int_{-\pi}^\pi g(\theta)K(|\theta -\phi|)
h(\phi)){\rm d}\theta{\rm d}\phi \Eqa(A.11)$$
for all ${\bf u}$. But according to Lemma 5.3, this is only possible if,
up to a common rotation,  $g = g^\star$ and $h=h_\star$.
Now repeat the argument for each of the coordinates. A similar argument
applies to  \eqv(reran2).  \eop

\noindent{\bf Proof of Lemma A.1} Given a unit vector ${\bf u}$, let
${{\cal S}_+}$ denote the set of points $x$ in ${\cal S}^1$ such that
$$({\bf x}\cdot {\bf u})({\bf u}\cdot {\bf u_0})\ge 0\ ,\Eqa(A.12)$$
where as before, ${\bf u}_0$ is the ``north pole'' $(0,1)$.
Let ${\cal S}_-$ be the complement of ${{\cal S}_+}$. That is, the line
fixed by $R_{{\bf u}}$ slices $S^1$
in two, and ${{\cal S}_+}$ is the half containing the north pole.
Then since $R_{{\bf u}}$ is a measure preserving transformation,
$$\eqalign{
&\int_{-\pi}^\pi \int_{-\pi}^\pi g(\theta)K(|\theta -\phi|)h(\phi) {\rm
d}\theta{\rm d}
\phi =\cr &\int_{{\cal S}_+}\int_{{\cal S}_+} g(\theta)K(|\theta
-\phi|)h(\phi) {\rm
d}\theta{\rm d}\phi +
\int_{{\cal S}_-}\int_{{\cal S}_-} g(\theta)K(|\theta -\phi|)h(\phi)
{\rm d}\theta{\rm d}
\phi +\cr &\int_{{\cal S}_-}\int_{{\cal S}_+} g(\theta)K(|\theta
-\phi|)h(\phi) {\rm
d}\theta{\rm d}\phi +
\int_{{\cal S}_+}\int_{{\cal S}_-} g(\theta)K(|\theta -\phi|)h(\phi)
{\rm d}\theta{\rm d}
\phi =\cr &\int_{{\cal S}_+}\int_{{\cal S}_+} g(\theta)K(|\theta
-\phi|)h(\phi) {\rm
d}\theta{\rm d}\phi +
\int_{{\cal S}_+}\int_{{\cal S}_+} g(R_{\bf u}\theta)K(|\theta
-\phi|)h(R_{\bf u}\phi)
{\rm d}\theta{\rm d}\phi\cr &\int_{{\cal S}_+}\int_{{\cal S}_+} g(R_{\bf
u}\theta)K(|R_{\bf
u}\theta -\phi|)h(\phi) {\rm d}\theta{\rm d}\phi +
\int_{{\cal S}_+}\int_{{\cal S}_+} g(\theta)K(|\theta-R_{\bf u}\phi|)
h(R_{\bf
u}\phi) {\rm d}\theta{\rm d}\phi
\cr\ .}\Eqa(A.13)$$

The desired inequality is then a consequence of the following inequality
for paris of real
numbers: Let $a_1$ and $a_2$ and $b_1$ and $b_2$ be any for positive
real numbers.
Rearrange $a_1$ and $a_2$ to decrease, and $b_1$ and $b_2$ to increase;
i.e., let Let
$a^\star_1 = \max\{a_1,a_2\}$ and let $a^\star_2 = \min\{a_1,a_2\}$, and let
$b^\star_1 = \min\{b_1,b_2\}$ and let $b^\star_2 = \max\{b_1,b_2\}$. Then
$$a_1^\star b_1^\star + a_2^\star b_2^\star \le a_1b_1 + a_2 b_2\
,\Eqa(hardy)$$
and there is equality if and only if $a_1 = a_1^\star$ and $b_1 =
b_1^\star$ or
$a_1^\star=a_2$, $b_1^\star=b_2$.

We now apply this with
$$a_1 = g(\theta )\quad a_2 = g(R_{\bf u}\theta)\quad b_1 =
h(\phi)\quad{\rm and} \quad
b_2 = h(R_{\bf u}\phi)\ .\Eqa(A.15)$$ Then
$$a_1^\star = R_{\bf u}^+g(\theta )\quad a_2^\star = R_{\bf u}^+g(R_{\bf
u}\theta )
\quad b_1^\star = R_{\bf u}^-h(\phi)\quad{\rm and}\quad b_2^\star =
R_{\bf u}^-h(R_{\bf u}\phi)\ .\Eqa(A.16)$$
Since $$K(|\theta -R_{\bf u}\phi|) = K(|R_{\bf u}\theta -\phi|) <
K(|\theta -\phi|)\Eqa(A.17)$$
almost everywhere, we have that
$$\eqalign{
&g(\theta)K(|\theta -\phi|)h(\phi) +
g(R_{\bf u}\theta)K(|\theta -\phi|)h(R_{\bf u}\phi) +\cr
&g(\theta)K(|R_{\bf u}\theta -\phi|)h(\phi) +
g(R_{\bf u}\theta)K(|R_{\bf u}\theta -\phi|)R_{\bf u}^-h(R_{\bf u}\phi)
\ge\cr
&R_{\bf u}^+g(\theta)K(|\theta -\phi|)R_{\bf u}^-h(\phi) +
R_{\bf u}^+g(R_{\bf u}\theta)K(|\theta -\phi|)R_{\bf u}^-h(R_{\bf
u}\phi) +\cr
&R_{\bf u}^+g(\theta)K(|R_{\bf u}\theta -\phi|)R_{\bf u}^-h(\phi) +
R_{\bf u}^+g(R_{\bf u}\theta)K(|R_{\bf u}\theta -\phi|)R_{\bf
u}^-h(R_{\bf u}\phi)\cr}\Eqa(A.18)$$
for almost every $\theta$ and $\phi$ in ${\cal S}_+$, with equality if
and only if
$$g(R_{\bf u}\theta) \le g(\theta)\qquad{\rm and}\qquad h(R_{\bf u}\phi)
\ge g(\phi)\Eqa(AA11)$$
or
$$g(R_{\bf u}\theta) \ge g(\theta)\qquad{\rm and}\qquad h(R_{\bf u}\phi)
\le g(\phi)\Eqa(AA22)$$
for almost every $\theta$ and $\phi$ in ${\cal S}_+$. Now unless $g$ is
constant, we can find a
$\theta$ and $u$ so that either $g(R_{\bf u}\theta) < g(\theta)$ or
$g(R_{\bf u}\theta) > g(\theta)$.
Suppose it is the first case. Then \eqv(AA11) holds, and
for almost every $\phi$, we must have $h(R_{\bf u}\phi)
\ge g(\phi)$. Making a similar argument for $h$, we see that one of
\eqv(AA11) or \eqv(AA22)
must hold for almost every $\theta$ and $\phi$.
The only way that
this can happen is if $g$ and $h$
are symmetric monotone.

The key here is the following pointwise inequality:
If $a >b$ and $c > d$, then
$$F(a,d)+F(b,c) < F(a,c)+F(b,d)\ .\Eqa(cheddar2)$$
To see this, let $k = c-d$ and $h = a-b$. Then
$$\left(F(a,d)+F(b,c)\right) - \left(F(a,c)+F(b,d)\right) =
-hk\int_0^1\int_0^1{\partial^2 F\over
\partial \r_1\partial\r_2}(b+sh,d+tk){\rm d}s{\rm d}t\ .\Eqa(A.22)$$
The second inequality of Lemma A.1 follows directly from this.
\quad\eop
\medskip
\noindent{\bf Remark} As can be seen from the proof, the decrease upon
rearrangement can be estimated quantitatively if we
strengthen \eqv(cheddar1) so that the positive lower bound is uniform.
This could be used to relax the conditions on the
interactions.
\bigskip\bigskip\bigskip
\noindent{\bf Appendix B.}
\numfor= 1
\bigskip
In this appendix we prove the following
theorem which  extends Theorem 2.3 to the case where no a priori bound 
on the
densities is assumed, or the entropy term does not necessarily prevents 
vacuum, or both.
\vskip .3cm
\noindent{\bf Theorem B.1} {\it Assume:
\item{i)}$G$ and $D$  strictly convex functions
\item{ii)} $\lim_{r\to \infty} {F(r)\over r} = \infty\ ,$
\item{iii)} $G$ and $D$ satisfy the following  {\it doubling condition}:
there exists a constant $K$ so that for all $r$,
$$G(2r) \le KG(r)\qquad{\rm and}\qquad D(2r) \le KD(r)\ .\Eqb(double)$$
\item{iv)} there is an $L>-\infty$ so that
for all $r\ge 0$,
$$G(r)\ge L \qquad{\rm and}\qquad D(r) \ge L\ .\Eqb(2)$$
Then, under the same assumptions on the interactions of Theorem 2.3, the 
conclusions of Theorem 2.3 are still
true.}
\medskip

\medskip
\noindent{\bf Proof:} We first observe that for any fixed
admissible
$\ro^{(0)}$, the
functional ${\cal G}(\rho_2)$ defined by
$${\cal G}(\rho_2) =  \F(\ro^{(0)},\rt)\Eqb(GGdef)$$
is strictly convex on the set of densities satisfying the mean density
constraint. By Fatou's lemma, it is also lower semicontinuous in the $L^1$
topology.

Now since we seek minimizers of $\F$, we may assume that $\ro^{(0)}$ is
symmetric monotone about the point
on the torus that is antipodal to the origin, and then we may restrict our
consideration of
$\rt \mapsto \F(\ro^{(0)},\rt)$ to densities that are symmetric monotone
about
the origin.

Now for any fixed constant $M$, the set of densities $\rt$ such that $\rt(x)
\le
M$ for all $x$, and that $\rt$ is symmetric monotone
is strongly compact in $L^1(\La)$ by the Helly selection principle. It
therefore
follows that there is a unique minimizer
$\tilde \rt$ of $\rt \mapsto \F(\ro^{(0)},\rt)$ in this class. By Lemma 2.2,
this is also the unique minimizer of
${\cal G}(\rt)$ in the class of densities satisfying:

\medskip
\item{(i)} $\int _{\Lambda}\rt {\rm d}x = n_2$.
\medskip
\item{(ii)} $\rt$ is symmetric monotone with its maximum at the point
antipodal
to the origin.
\medskip
\item{(iii)} $\|\rt\|_\infty \le M$

We now show that for $M$ large enough,
this minimizer ceases to
depend on $M$, so that the condition (iii) becomes superfluous.

Consider a variation $\tilde \rt + h$ of $\tilde \rt$. Clearly, we must have
$h
\le 0$ on
$$A_M = \{ x\ | \tilde \rt(x)= M\}\ ,\Eqb(3)$$ and
$h\ge 0$ on
$$A_Z = \{ x\ | \tilde \rt(x)= 0\}\ .$$ We must also have
$$\int_{\La}h(x){\rm d}x = 0\Eqb(admis)$$
in order to preserve the condition that $(\ro^{(0)},\tilde \rt + h)$
belongs to
${\cal D}(n_1,n_2)$.
The Euler--Lagrange condition then is
$$\int_{\La}\left(F'(\tilde \rt) + D'(\ro^{(0)} + \tilde \rt) +
U*\ro^{(0)}\right)h{\rm d}x \ge 0\Eqb(elineq)$$

Let $B$ denote $(A_M\cup A_Z)^c$. We first show that for $M$ sufficiently
large,
$|B| \ne 0$.
Indeed, if $|B| = 0$, then $\tilde \rho_2 = M1_{A_M}$. Since
$\int_{\Lambda}\tilde \rho_2(x){\rm d}x =
n_1|\Lambda|$, $|A_Z|$ then equals $n_2|\Lambda|/M$ and
consequently,
$${\cal G}(\tilde \rho_2) \ge {F(M)\over M}n_2|\Lambda|\ .\Eqb(7)$$
Since $F(M)/M$ tends to infinity with $M$, and since ${\cal G}(\tilde
\rho_2)
\le {\cal G}(n_2/|\Lambda|)$
whenever $M \ge n_2/|\Lambda|$ so that $\rho_2 = n_2/|\Lambda|$ is an
admissible
trial density,
we obtain a contradiction. Henceforth take $M$ large enough to ensure that
$|B|
\ne 0$.
Note that on $B$
that the Euler Lagrange equation
$$F'(\tilde \rt) + D'(\ro^{(0)} + \tilde \rt) + U*\ro^{(0)} = C\Eqb(ELEL)$$
holds for some value of $C$, since on this set, if $h$ is an admissible
variations, so is $-h$.

We have  the Chebyshev estimate
$$|A_Z| \le {n_2|\Lambda|\over M}\ .\Eqb(9)$$ Further increasing $M$, we
may assume that $|A_Z| < |\Lambda|/3$.
It then follows that
$${ \rm either}\qquad |B| \ge {|\Lambda|\over 3}\qquad{\rm or}\qquad
  |A_Z| \ge {|\Lambda|\over 3}\ .\Eqb(10)$$
Our next goal is to obtain an {\it a--priori} bound on $C$. We will obtain
two
such bounds: one that is valid,
when $|B|$ is not too small, and one that is valid when $|A_Z|$ is not too
small. By the above, at least one of
these two must hold.

To obtain a bound that will be useful if $|B|$ is not too small,
integrate the Euler Lagrange equation over $B$, and obtain
$$\int_B F'(\tilde \rho_2){\rm d}x +
\int_B D'(\rho_1^{(0)} +\tilde \rho_2){\rm d}x + \int_B U*\rho_1^{(0)}{\rm
d}x =
C|B|\ .\Eqb(11)$$
Now, by convexity, for any $a$,
$$F'(a) \le F(a+1) - F(a) \le F(a+1) = F\left({2a+ 2\over 2}\right) \le
{1\over
2}F(2a) + {1\over 2}F(2)\ .\Eqb(12)$$
Making use of the doubling condition, we finally have
$$F'(a)  \le {K\over 2}F(a) + {1\over 2}F(2)\ .$$
Replacing $a$ by $\tilde \rho_2$ and integrating over $B$,
$$\int_B F'(\tilde \rho_2){\rm d}x \le {K\over 2}\int_{\Lambda} F(\tilde
\rho_2){\rm d}x +
{1\over 2}F(2)|\lambda|\ .\Eqb(13)$$
In the same way, we obtain
$$\int_B D'(\rho_1^{(0)} +\tilde \rho_2){\rm d}x \le
{K\over 2}\int_{\Lambda}D(\rho_1^{(0)} +\tilde \rho_2){\rm d}x + {1\over
2}D(2)|\lambda|\ .\Eqb(14)$$
Finally,
$$\int_B U*\rho_1^{(0)}{\rm d}x \le \alpha |\Lambda|n_1\ .\Eqb(15)$$
Combining estimates we have
$$C \le {1\over |B|}\left({K\over 2}{\cal G}(n_2/|\Lambda|) + {1\over
2}\left(F(2) + D(2)\right) + \alpha
n_1|\Lambda|\right)\ ,\Eqb(16)$$
so that under the first alternative above, we have the {\it a--priori} bound
$$C \le
{3\over |\Lambda|}\left({K\over 2}{\cal G}(n_2/|\Lambda|) + {1\over
2}\left(F(2)
+ D(2)\right) + \alpha
n_1|\Lambda|\right)\ .\Eqb(17)$$

Next, we obtain an estimate that will be useful is $|A_Z|$ is not too
small.
First, if $|A_Z| \ne 0$, we can consider a variation of the following type:
Let $h$ satisfying \eqv(admis) with $h_+$
supported by $A_Z$, and with $h_-$ supported in
$B$. Then if we let $a = \int_{\La}h_+{\rm d}x$, we have from \eqv(elineq)
that
$$F'(0) + {1\over a}\int_{\La}\left( D'(\ro^{(0)}) +
U*\ro^{(0)}\right)h_+{\rm
d}x \ge C\ .\Eqb(EL2)$$

By this variational inequality,
$$C \le F'(0) + {1\over |A_Z|}\int_{A_Z}\left( D'(\rho_1^{(0)}) +
U*\rho_1^{(0)}\right){\rm d}x\ .\Eqb(19)$$
The same convexity arguments employed above now yield
$$C \le  F'(0) + {1\over |A_Z|}\left(
{K\over 2}{\cal G}(n_2/|\Lambda|) + {1\over 2} D(2) + \alpha
n_1|\Lambda|\right)\ ,\Eqb(20)$$
which, under the second alternative above, becomes
$$C \le F'(0) +
{3\over |\Lambda|}\left(
{K\over 2}{\cal G}(n_2/|\Lambda|) + {1\over 2} D(2) + \alpha
n_1|\Lambda|\right)\ .\Eqb(21)$$
Now notice that in the case $F(t) = t\log t$, $F'(0) = -\infty$, so in this
case, $|A_Z| > 0$ is precluded.
We therefore have the {\it a--priori} estimate
$$C \le \max\{F'(0),0\} +
{3\over |\Lambda|}\left({K\over 2}{\cal G}(n_2/|\Lambda|) + {1\over
2}\left(F(2)
+ D(2)\right) + \alpha
n_1|\Lambda|\right)\ .\Eqb(22)$$

To apply his, we suppose $|A_M|>0$, and consider a variation of the
following
type:
Let $h$ satisfy \eqv(admis) with $h_-$, the negative part of $h$,
supported by $A_M$, and with $h_+$ supported in
$B$. Then if we let $a = \int_{\La}h_+{\rm d}x$, we have from \eqv(elineq)
that
$$-\int_{\La}\left(F'(\tilde \rt) + D'(\ro^{(0)} + \tilde \rt) +
U*\ro^{(0)}\right)h_-{\rm d}x  + Ca \ge 0\Eqb(23)$$
This means that
$${1\over a}\int_{\La}\left(F'(\tilde \rt) + D'(\tilde \rt) +
U*\ro^{(0)}\right)h_-{\rm d}x \le C\ ,\Eqb(EL12)$$
which in turn, since $D'$ is an increasing function,  means that
$$F'(M)  \le C\ .\Eqb(EL3)$$
Let $G$ be the inverse function to $F'$;
i.e.,
$$G(t) = \inf\{ r \ge 0 \ | \ F'(r) \ge t\ \}\ .\Eqb(26)$$
Then, we have the {\it a--priori} bound
$$M \le G(C)\ ,$$
which, when combined with our {\it a--priori} bound on $C$ implies that for
sufficiently large $M$,
$|A_M| = 0$.

Therefore, our minimizer $\tilde \rho_2$ satisfies
$$\tilde\rho_2 \le
G\left(\max\{F'(0),0\} +
{3\over |\Lambda|}\left({K\over 2}{\cal G}(n_2/|\Lambda|) + {1\over
2}\left(F(2)
+ D(2)\right) + \alpha
n_1|\Lambda|\right)\right)\ .\Eqb(27)$$
Moreover, we see that wherever $\tilde \rho_2 \ne 0$, it satisfies the
Euler--Lagrange equation.

Now fix this minimizer $\tilde \rho_2$ and consider the functional
$${\cal G}_1 (\rho_1) = {\cal F}(\rho_1,\tilde \rho_2)\ .\Eqb(28)$$
Exactly the same argument shows that this has a minimizer $\tilde \rho_1$ in
the
class of densities that are
symmetric monotone with a maximum at the origin, and with
$\int_{\Lambda}\rho_1{\rm d}x = |\Lambda|n_1$.
Moreover, we obtain in this way the {\it a--priori} bound
$$\tilde\rho_1 \le
G\left(\max\{F'(0),0\} +
{3\over |\Lambda|}\left({K\over 2}{\cal G}(n_1/|\Lambda|) + {1\over
2}\left(F(2)
+ D(2)\right) + \alpha
n_2|\Lambda|\right)\right)\ .\Eqb(29)$$

Now consider a minimizing sequence $(\rho_1^{(k)}, \rho_2^{(k)})$ in ${\cal
D}(n_1,n_2)$ for ${\cal F}$.
By the argument above, we can replace each pair
$(\rho_1^{(k)}, \rho_2^{(k)})$ by another pair $(\tilde \rho_1^{(k)},
\tilde\rho_2^{(k)})$ in
${\cal D}(n_1,n_2)$ so that $\tilde \rho_1^{(k)}$ and $\tilde\rho_2^{(k)}$
are
symmetric monotone about the origin
and its antipodal point respectively, and so that $\tilde \rho_1^{(k)}$ and
$\tilde\rho_2^{(k)}$
satisfy the {\it a--priori} $L^\infty$ bounds above, and last but not least,
so
that
$${\cal F}(\tilde \rho_1^{(k)}, \tilde\rho_2^{(k)}) \le {\cal
F}(\rho_1^{(k)},
\rho_2^{(k)})\ .\Eqb(30)$$
Now by the Helly selection principle again, we have that for a subsequence,
$$\tilde \rho_1  = \lim_{n\to \infty}\tilde \rho_1^{(k_n)}\Eqb(31)$$
and
$$\tilde \rho_2  = \lim_{n\to \infty}\tilde \rho_2^{(k_n)}\Eqb(32)$$
exist almost everywhere. By the {\it a--priori} $L^\infty$ bounds, and the
dominated convergence theorem,
these limits also hold in $L^1$, and hence $(\tilde \rho_1,\tilde \rho_2)
\in
{\cal D}(n_1,n_2)$. Moreover, by the lower semicontinuity of ${\cal F}$
discussed at the outset,
$${\cal F}(\tilde \rho_1,\tilde \rho_2) \le \liminf_{n\to \infty}
{\cal F}(\tilde \rho_1^{(k_n)}, \tilde\rho_2^{(k_n)}) =
\inf_{(\rho_1,\rho_2)\in
{\cal D}(n_1,n_2)}
{\cal F}(\rho_1,\rho_2)\ .\Eqb(33)$$
This proves the existence of  our minimizers.

We now drop the tilde, and examine their properties. By the argument above,
we
clearly have
$$\rho_1(x) = G\left(C_1 - U*\rho_2(x) - D'(\rho_1(x) +
\rho_2(x))\right)\Eqb(34)$$
on the set where $\rho_1(x) \ne 0$, and
likewise
$$\rho_2(x) = G\left(C_2 - U*\rho_1(x) - D'(\rho_1(x) +
\rho_2(x))\right)\Eqb(35)$$
on the set where $\rho_2(x) \ne 0$, and we
have the asserted $L^\infty$ bounds.
Finally, since
$\rho_2$ and $\rho_1$ are monotone symmetric, these $L^\infty$ bounds imply
$L^1$ bounds for the
gradients of $\rho_1$ and $\rho_2$. \eop
\medskip

{\leftskip 1cm\rightskip1cm
\vskip1cm\centerline{\bf References}\vskip1cm

\noindent{[\rtag{ABCP}]}  Alberti G., Bellettini G.,  Cassandro M. and 
Presutti  E., { Surface
Tension in Ising Systems with kac Potentials}, J.Stat.Phys, {\bf 82}, 
743--796 (1996).

\noindent{[\rtag{AL}]} Almgren  F. and  Lieb E.,
{ Symmetric rearrangement is sometimes continuous},
J. Amer. Math. Socs., {\bf 2}, 683--733 (1989).
\vskip.1cm

\noindent{[\rtag{BT}]}  Baernstein A. and  Taylor B.,
{ Spherical rearrangements, subharmonic
functions, and $*$--functions in $n$--space},
Duke Math. Jour. {\bf 43}, 245--268 (1976).
\vskip.1cm

\noindent{[\rtag{BELM}]}  Bastea S.,  Esposito R., Lebowitz  J. L. and
Marra R., {
Binary Fluids with
Long Range Segregating Interaction I:
Derivation of Kinetic and Hydrodynamic Equations,}  J. Stat.
Phys., {\bf 101}, 1087--1136 (2000).
\vskip.1cm
\noindent
{[\rtag{db}]} J. de Boer, Van der Waals
in His Time and The Present Revival - Opening Address, in Van der Waals
Centennial Conference on Statistical Mechanics, Amsterdam, 27-31 August
1973, C. Prins, editor, North-Holland, Amsterdam, 1-17, 1974.
\vskip.1cm
\noindent{[\rtag{DOPT}]}
De Masi A., Orlandi E., Triolo L. and Presutti E.
{ Glauber evolution with Kac potentials I. Mesoscopic and macroscopic 
limits, interface
dynamics, } Nonlinearity {\bf 7}, 287--301, (1996).
{ Glauber evolution with Kac potentials II. Fluctuation. }
  Nonlinearity {\bf 9}, 27--51, (1996).
{ Glauber evolution with Kac potentials. III. Spinodal decomposition. } 
Nonlinearity {\bf 9},
53--114, (1996).
\vskip.1cm
\noindent{[\rtag{Ge}]}   Georgi H.O., {\it Gibbs measures and phase 
transitions}, W. De. Gruyter, Berlin (1988).
\vskip.1cm
\noindent{[\rtag{GP}]} Gates D.J. and Penrose O., { The van der Waals 
limit for classical systems.
I. A variational principle.} Commun. Math. phys. {\bf 15} 255--276 (1969).
\vskip.1cm
\noindent{[\rtag{KUH}]}  Kac M.,  Uhlenbeck G.,  Hemmer P.C.
{\  On the Van der Waals theory of
vapor-liquid equilibrium }{ I. Discussion of
a one dimensional model},  J. Math. Phys.
   {\bf 4},
     216--228
      (1963);
{\ II.  Discussion of the distribution
functions }, J. Math. Phys., {\bf 4},  229--247,
     (1963);
{ III. Discussion of the critical
region}  J. Math. Phys., {\bf 5}.
     60--74,
     (1964)
    \vskip.1cm
     \noindent{[\rtag{K}]}  Korteweg D.J., {Sur le Points de 
Plissements} Arch. n\'eerl. 24
(1891) 57--98; { Theorie g\'en\'erale des Plis}, Arch. N\'eerl. 24 
295--368 (1891).
\vskip.1cm
\noindent{[\rtag{vK}]}  Van Kampen N.G., { Condensation of a classical 
gas with long-range
attraction}, Phys Rev. A {\bf 135} 362--369 (1964).
  \vskip.1cm
\noindent{[\rtag{LP}]}  Lebowitz J.L. , Penrose  O.,
{  Rigorous treatment of the Van der Waals
Maxwell theory of the liquid vapor transition}
   J. Math. Phys., {\bf 7},  98,
     (1966)
     \vskip.1cm
    \noindent{[\rtag{LMP}]}  Mazel A.,   Lebowitz J. L. and  Presutti 
E., { Liquid-vapor phase
transitions for systems with finite-range interactions}  J. Statist. 
Phys.  {\bf 94},  no. 5-6,
955--1025 (1999).
\vskip.1cm
\noindent{[\rtag{L}]} Luttinger J. ,
{ Generalized isoperimetric inequalities},
J. Math. Phys., {\bf 14}, 1448--1450 (1973).
\vskip.1cm
\noindent{[\rtag{PS}]} Pirogov S.A. and Sinai G.Ya, Phase diagrams of 
classical lattice systems.
Teoret. Mat. Fiz.  25,
  358--369 (1975).
\vskip.1cm
\noindent{[\rtag{Ro}]} J. Rowlingson; J. D. van der Waals: On the Continuity of
the Gaseous and Liquid States, North Holland (1988).
\vskip.1cm
\noindent{[\rtag{Ru}]}  Ruelle D., {\it Statistical Mechanics: Rigorous 
Results}. Benjamin, New York
(1969);
Some Ill-Formulated Problems on
Regular and Messy Behavior in Statistical Mechanics and Smooth Dynamics for
which I Would Like the Advice of Yasha Sinai, J. Stat. Phys., {\bf 108}, 72--729,
2002.
\vskip.1cm
\noindent{[\rtag{LS}]}  Levelt Sengers J.M.H., {\it How Fluids Unmix; 
Discoveries by the School of Van
der Waals and Kamerlingh Onnes} (Edita, Nederlands).
\vskip.1cm
\noindent{[\rtag{Sim}]}  Simon B., {\it The  Statistical Mechanics of 
Lattice Gases}, Vol I.
Princeton University Press, Princeton, New Jersey (1993).
 \vskip.1cm
\noindent{[\rtag{Si}]}  Sinai G. Ya., {\it Theory of phase transitions: 
rigorous results}.
International Series in Natural Philosophy, 108. Pergamon Press, 
Oxford-Elmsford, N.Y., 1982.
\vskip.1cm
\noindent{[\rtag{VW}]}  van der Waals J. D., I Verhandelingen, 
Kon.Akad. Wet. Amsterdam 20 (1880);
II{ Th\'eorie mol\'eculaire d'une substance compos\'ee de deux 
mati\`eres diff\'erentes }, Arch.
n\'eerl. 24, 1  (1891).
\vskip.1cm
\noindent{[\rtag{Za}]} M. Zahradnik, An Alternate Version of Pirogov-Sinai
Theory, Comm. Math. Phys. {\bf 93}, 559-581 (1984).  
\end